\documentclass[fleqn,twocolumn,trackchanges]{aastex701}
\usepackage{mathtools}  
\usepackage{nccmath}    
\usepackage{braket}   
\usepackage{booktabs}
\usepackage{graphicx}
\usepackage{natbib}  
\usepackage{physics}  
\usepackage[utf8]{inputenc} 
\usepackage{tikz}
\usepackage{tikz-3dplot}

\begin{document}

\title{Predicting Primordial Non-Gaussianity Measurements in the CSST Spectroscopic Survey}

\correspondingauthor{Yan Gong}

\author{Jiangnan Duan}
\affiliation{National Astronomical Observatories, Chinese Academy of Sciences,20A Datun Road, Beijing 100012, China}
\affiliation{School of Astronomy and Space Sciences, University of Chinese Academy of Sciences(UCAS),\\Yuquan Road NO.19A Beijing 100049, China}
\email{duanjn@bao.ac.cn}  

\author{Yan Gong} 
\affiliation{National Astronomical Observatories, Chinese Academy of Sciences,20A Datun Road, Beijing 100012, China}
\affiliation{School of Astronomy and Space Sciences, University of Chinese Academy of Sciences(UCAS),\\Yuquan Road NO.19A Beijing 100049, China}
\affiliation{Science Center for CSST, National Astronomical Observatories, CAS, 20A Datun Road, Beijing 100101, China}
\email[show]{gongyan@bao.ac.cn}

\author{Qi Xiong} 
\affiliation{National Astronomical Observatories, Chinese Academy of Sciences,20A Datun Road, Beijing 100012, China}
\affiliation{School of Astronomy and Space Sciences, University of Chinese Academy of Sciences(UCAS),\\Yuquan Road NO.19A Beijing 100049, China}
\email{}  

\author{Haitao Miao}
\affiliation{National Astronomical Observatories, Chinese Academy of Sciences,20A Datun Road, Beijing 100012, China}
\affiliation{State Key Laboratory of Radio Astronomy and Technology, National Astronomical Observatories, CAS,  \\A20 Datun Road,  Chaoyang District,  Beijing 100101,  People's Republic of China}
\email{}

\author{Xuelei Chen} 
\affiliation{National Astronomical Observatories, Chinese Academy of Sciences,20A Datun Road, Beijing 100012, China}
\affiliation{School of Astronomy and Space Sciences, University of Chinese Academy of Sciences(UCAS),\\Yuquan Road NO.19A Beijing 100049, China}
\affiliation{State Key Laboratory of Radio Astronomy and Technology,  National Astronomical Observatories, CAS,  \\A20 Datun Road,  Chaoyang District,  Beijing 100101,  People's Republic of China}
\affiliation{Department of Physics, College of Sciences, Northeastern University, Shenyang 110819, China}
\affiliation{Centre for High Energy Physics, Peking University, Beijing 100871, China}
\email{}  

\author{Qi Guo} 
\affiliation{Institute for Frontiers in Astronomy and Astrophysics, Beijing Normal University, Beijing 102206, China}
\affiliation{School of Physics and Astronomy, Beijing Normal University, Beijing 100875, China}
\affiliation{Key Laboratory for Computational Astrophysics, National Astronomical Observatories, \\Chinese Academy of Sciences, Beijing 100101, China}
\email{}  

\author{Ming Li} 
\affiliation{National Astronomical Observatories, Chinese Academy of Sciences,20A Datun Road, Beijing 100012, China}
\email{}  

\author{Yun Liu} 
\affiliation{National Astronomical Observatories, Chinese Academy of Sciences,20A Datun Road, Beijing 100012, China}
\affiliation{School of Astronomy and Space Sciences, University of Chinese Academy of Sciences(UCAS),\\Yuquan Road NO.19A Beijing 100049, China}
\email{}  

\author{Wenxiang Pei} 
\affiliation{Shanghai Key Lab for Astrophysics, Shanghai Normal University, Shanghai 200234, China}
\email{}  

\begin{abstract}

Primordial non-Gaussianity (PNG) is a fundamental probe of the physics of the early Universe and inflation. Here we present a comprehensive study of the constraints on the local-type PNG parameter, $f_{\rm NL}$, for the spectroscopic galaxy survey of the upcoming Chinese Space-station Survey Telescope (CSST). Utilizing the high-resolution Jiutian N-body simulation suite, we construct realistic mock catalogs for emission line galaxies (ELGs) at three representative redshifts $z=0.3$, 0.6, and 0.9. The expected CSST observational characteristics are also considered, including redshift uncertainties and selection functions based on signal-to-noise ratios of emission lines. We develop a robust analysis  framework for the redshift-space galaxy power spectrum and bispectrum that accounts for redshift-space distortions, scale-dependent bias, and nonlinear effects. {For the parameter $p$ in the PNG bias $b_{\phi}$ with $p=0.55$, we find  $f_{\rm NL}=-4\pm43$ if only using the power spectrum data, and $f_{\rm NL}=3\pm41$ after combining the bispectrum data, reducing the uncertainty by $\sim$5\%.} {For the $p=1$ case, the constraint results are $f_{\rm NL}=-3\pm86$ and $54^{+70}_{-80}$ for comparison.} 
This work demonstrates the potential of the Stage~IV surveys like CSST to probe inflationary physics, and highlights the importance of higher-order statistics in extracting information from large-scale structure surveys.

\end{abstract}


\section{Introduction} 

The measurements of primordial non-Gaussianity (PNG) is crucial for exploring the properties of the early Universe and testing different inflation models \citep{Maldacena_2003,Bartolo_2004,planckcollaboration2019planck2018resultsix}. At the level of the gravitational potential, the degree of deviation from Gaussianity can be quantified by the parameter \(f_{\mathrm{NL}}\). Its most common parameterization, corresponding to the ``local-type'' non-Gaussianity, is given by:
\begin{equation}
\label{eq:fNL_def}
\Phi(\mathbf{x}) = \varphi(\mathbf{x}) + f_{\mathrm{NL}} \left( \varphi^2(\mathbf{x}) - \langle \varphi^2(\mathbf{x}) \rangle \right).
\end{equation}
Here, \(\Phi(\mathbf{x})\) is the primordial gravitational potential field and \(\varphi(\mathbf{x})\) is an auxiliary Gaussian random field. In the standard single-field slow-roll inflation scenario, primordial density perturbations are predicted to follow nearly Gaussian statistics, with $f_{\mathrm{NL}} \sim \mathcal{O}(10^{-2})$ \citep{Maldacena_2003,Paolo_Creminelli_2004,Cabass_2017}. However, significant deviations from Gaussianity would provide decisive evidence for more sophisticated models of inflation, such as multi-field inflation that predicts $f_{\mathrm{NL}} \sim \mathcal{O}(1)$ ~\citep{Senatore_2012,alvarez2014testinginflationlargescale}.

 Currently, the tightest constraint on \(f_{\mathrm{NL}}\) comes from the analysis of the $\it Planck$ cosmic microwave background (CMB) bispectrum measurement: $f_{\rm NL}=-0.9\pm5.1$ \citep{planckcollaboration2019planck2018resultsix}. Since CMB observations are essentially a two-dimensional projection of the last scattering surface, the three-dimensional large-scale structure (LSS) measured by galaxy surveys offers a complementary and highly promising alternative approach. Galaxy surveys such as the Extended Baryon Oscillation Spectroscopic Survey (eBOSS) ~\citep{Cagliari_2025} and early data from the Dark Energy Spectroscopic Instrument (DESI) ~\citep{chaussidon2025} are providing increasingly tighter constraints of PNG through power spectrum and bispectrum analysis. The next generation of galaxy redshift survey projects, including the  Chinese Space-station Survey Telescope (CSST) ~\citep{132011-961,TB-2021-0016,Gong_2019,Gong_2025,2026},  will cover unprecedented volumes with high spectroscopic completeness, promising to significantly improve PNG constraints.

This work presents a comprehensive study of the PNG constraining power of the CSST spectroscopic galaxy survey by using high-resolution cosmological numerical simulations. {We construct mock galaxy catalogs that incorporate selected observational characteristics of CSST and systematically develop analysis pipelines for both the galaxy power spectrum and the bispectrum.} {The analysis uses plane-parallel cuboid mocks and does not include the actual CSST angular footprint or its associated survey-window convolution.} {The bispectrum provides configuration-dependent information on local PNG, particularly through the coupling between long- and short-wavelength modes in squeezed configurations.} {However, it also contains contributions from late-time gravitational evolution, nonlinear galaxy bias, and redshift-space distortions, which must be modeled jointly \citep{Tellarini_2016}.} {We therefore investigate how much additional information the selected bispectrum configurations provide beyond the power spectrum in our CSST-like mock samples.}

This paper is organized as follows:  Section~\ref{sec:theory} elaborates on the theoretical models for the galaxy power spectrum and bispectrum. Section~\ref{sec:mock_data} describes the construction of mock galaxy catalogs and the estimation of power spectrum and bispectrum. In Section~\ref{sec:results} we present the constraint results on \(f_{\rm NL}\)  from the mock data and analyze the contributions from different statistics and redshift samples. We summarize our results in Section~\ref{sec:conclusion}.

\section{Theoretical Models} \label{sec:theory}

We model the galaxy power spectrum and bispectrum by incorporating the effects of local-type PNG, non-linear biasing, and redshift-space distortions (RSD) in our theoretical framework \citep{Tellarini_2016}.

\subsection{Galaxy Power Spectrum}

In redshift space, the galaxy power spectrum is modeled by considering the linear Kaiser effect, the scale-dependent bias induced by $  f_{\rm NL}$ , and a Lorentzian-type damping term representing the ``Fingers-of-God" (FoG) effect. The 2D power spectrum $P_g(k, \mu)$ is given by
\begin{equation}
P_g(k, \mu) = \frac{\left[ {b_1} + f \mu^2 + f_{\rm NL} \frac{b_\phi}{M(k)} \right]^2}{\left[ 1 + \frac{1}{2}(k \mu \sigma_P)^2 \right]^2} P_L(k),
\end{equation}
where $b_1$ is the linear bias of the tracer, $b_{\phi}$ is the PNG bias parameter, which encodes the tracer's response to the local-type PNG. $P_L(k)$ is the linear matter power spectrum, $f$ is the growth rate, and $\mu$ is the cosine of the angle between the wavevector $k$ and the line-of-sight. The term $M(k)$ relates the density field to the primordial potential via the Poisson equation, which is given by
\begin{equation}
M(k) = \frac{2c^2 k^2 T(k)}{3 \Omega_m H_0^2},
\end{equation}
where $T(k)$ is the matter transfer function normalized to unity at large scales. The parameter $\sigma_P$ is a free parameter accounts for the small-scale velocity dispersion. We use {\tt CLASS} \citep{Diego_Blas_2011} to calculate linear power spectrum \( P_{L}(k) \) and the transfer function $T(k)$.

The observable power spectrum multipoles $P_\ell(k)$ are obtained via the Legendre projection
\begin{equation}
P_\ell(k) = \frac{2\ell + 1}{2} \int_{-1}^{1} P_g(k, \mu) \mathcal{L}_\ell(\mu) d\mu.
\end{equation}
Here $\mathcal{L}_\ell(\mu)$ is the Legendre polynomials. In this analysis, we primarily focus on the monopole ($\ell=0$), which carry most of the information \citep[see e.g.][]{Cagliari_2025}.

\subsection{Galaxy Bispectrum}

The tree-level galaxy bispectrum in redshift space is modeled considering the second-order perturbation theory (2PT) kernels and PNG corrections, and the general form is given by
\begin{equation}
\begin{aligned}
 &B_g({\boldsymbol{k}_1, \boldsymbol{k}_2, \boldsymbol{k}_3}) = \mathcal{D}_{\text{FoG}}(\boldsymbol{k}_1, \boldsymbol{k}_2, \boldsymbol{k}_3)
\\
&\times2 \left[ Z_1(\boldsymbol{k}_1) Z_1(\boldsymbol{k}_2) Z_2(\boldsymbol{k}_1, \boldsymbol{k}_2) P_L(k_1) P_L(k_2) + {\text{2 cyc.}}\right], \label{eq:Bg}
\end{aligned}
\end{equation}
where ``2 cyc.'' denotes the addition of the two remaining cyclic permutations of the corresponding wavevectors. The first and second-order redshift-space kernels, $Z_1$ and $Z_2$, are modified by the presence of $f_{\rm NL}$:
\begin{itemize}
    \item \textbf{First-order kernel:}
    \begin{equation}
    Z_1(\boldsymbol{k}) = b_1 + f \mu^2 + f_{\rm NL} \frac{b_\phi}{M(k)};
    \end{equation}
    
    \item \textbf{Second-order kernel:}
    The $Z_2$ kernel includes contributions from non-linear clustering ($F_2$), non-linear RSD ($G_2$), and non-linear bias ($b_2, b_{s^2}, b_{\phi\delta}$):

\begin{equation}
\begin{split}
Z_2(\boldsymbol{k}_1, \boldsymbol{k}_2) &= \frac{b_2}{2} + b_{s^2} {s^2(\boldsymbol{k}_1, \boldsymbol{k}_2)} \\
&\quad + b_1 \left[ F_2(\boldsymbol{k}_1, \boldsymbol{k}_2) + \frac{f_{\rm NL} M(k_3)}{M(k_1)M(k_2)} \right] \\
&\quad + f \mu_3^2 \left[ G_2(\boldsymbol{k}_1, \boldsymbol{k}_2) + \frac{f_{\rm NL} M(k_3)}{M(k_1)M(k_2)} \right] \\
&\quad - \frac{f \mu_3 k_3}{2} \left[ \frac{\mu_1}{k_1} Z_1(\boldsymbol{k}_2) + \frac{\mu_2}{k_2} Z_1(\boldsymbol{k}_1) \right] \\
&\quad + \frac{b_\phi f_{\rm NL}}{2} \left[ \frac{k_1/k_2}{M(k_1)} + \frac{k_2/k_1}{M(k_2)} \right] \mu_{12} \\
&\quad + \frac{b_{\phi\delta} f_{\rm NL}}{2} \left[ \frac{1}{M(k_1)} + \frac{1}{M(k_2)} \right],
\end{split}
\end{equation}

\end{itemize}
where $s^2$ is the traceless part of the shear field, \( F_2 \) and \( G_2 \) are the second-order matter density and velocity kernels, and \( b_2 \) and \( b_{s^2} \) are the quadratic local bias and the second-order tidal bias, respectively. We have 
\begin{equation}
\begin{aligned}
{s^{2}(\boldsymbol{k}_{1},\boldsymbol{k}_{2})} &= \left(\hat{\boldsymbol{k}}_{1}\cdot\hat{\boldsymbol{k}}_{2}\right)^{2} - \frac{1}{3},
\end{aligned}
\end{equation}
\begin{equation}
\begin{aligned}
F_{2}(\boldsymbol{k}_{1},\boldsymbol{k}_{2}) &= \frac{5}{7} + \frac{1}{2}\left(\frac{k_{1}}{k_{2}} + \frac{k_{2}}{k_{1}}\right)\mu_{12} + \frac{2}{7}\mu_{12}^{2},
\end{aligned}
\end{equation}
\begin{equation}
\begin{aligned}
G_{2}(\boldsymbol{k}_{1},\boldsymbol{k}_{2}) &= \frac{3}{7} + \frac{1}{2}\left(\frac{k_{1}}{k_{2}} + \frac{k_{2}}{k_{1}}\right)\mu_{12} + \frac{4}{7}\mu_{12}^{2}.
\end{aligned}
\end{equation}
To reduce the number of free parameters, we adopt relations derived from the Lagrangian bias model and the peak background split (PBS) formalism:
\begin{itemize}
    \item \textbf{Tidal Bias:} We assume the local-in-Lagrangian-space constraint~\citep{Chan_2012}:
    \begin{equation}
    b_{s^2} = -\frac{2}{7}(b_1 - 1).
    \end{equation}
    
    \item \textbf{PNG Bias ($b_\phi$):} Following the PBS relations for halo mass functions \citep{Baldauf_2011}: 
    \begin{equation}
    b_\phi = 2 \delta_c (b_1 - p),
    \end{equation}
where \( \delta_c \approx 1.686 \) is the spherical-collapse threshold and \(p\) parametrizes the response of the selected tracer population. {In our analysis, we explore the results for $p=1$ and $p=0.55$ cases to investigate how the inferred PNG constraints depend on the assumed tracer response. 
The choice $p=1$ recovers the widely used universality relation, corresponding to the conventional universal-mass-function prediction $b_\phi=2\delta_c(b_1-1)$, as a standard reference.
For \(p\simeq0.55\), it has been found to better describe stellar-mass-selected galaxies \citep{Barreira_2022}. In our simulation, emission-line luminosities are linked to star formation activity \citep{Pei2024}, which correlates with stellar mass along the star-forming main sequence \citep{Noeske2007}. This connection motivates adopting $p=0.55$ as an approximate fiducial choice for our SNR-selected galaxies.}
    
    \item \textbf{Non-linear PNG Bias ($b_{\phi\delta}$):} We use the relation from \cite{Baldauf_2011}: 
    \begin{equation}
    b_{\phi\delta} = b_\phi + 2 \left[ \delta_c \left( b_2 - \frac{8}{21}(b_1 - 1) \right) - b_1 + 1 \right].
    \end{equation}
\end{itemize}
To account for small-scale damping in the bispectrum, we apply a FoG factor $\mathcal{D}_{\text{FoG}}$, which can be expressed as
\begin{equation}
\mathcal{D}_{\text{FoG}} = \left[ 1 + \frac{\sigma_B^4}{2} (k_1^2 \mu_1^2 + k_2^2 \mu_2^2 + k_3^2 \mu_3^2)^2 \right]^{-2}.
\end{equation}
The set of free parameters in our analysis is $\{b_1, b_2, f_{\rm NL}, \sigma_P, \sigma_B\}$.

\begin{figure}[t]
    \centering
    \tdplotsetmaincoords{70}{115} 
    \begin{tikzpicture}[tdplot_main_coords, scale=3.5, >=stealth]
        
        \draw[->] (0,0,0) -- (1.2,0,0) node[anchor=north east]{$\hat{x}$};
        \draw[->] (0,0,0) -- (0,1.2,0) node[anchor=north west]{$\hat{y}$};
        \draw[thick, ->] (0,0,0) -- (0,0,1.3) node[anchor=south]{$\hat{z} \, (\hat{n})$};

        \pgfmathsetmacro{\thetaOne}{30}
        \pgfmathsetmacro{\phiOne}{-20}
        \pgfmathsetmacro{\thetaTwo}{30}
        \pgfmathsetmacro{\phiTwo}{110}

        \tdplotsetcoord{K1}{1}{\thetaOne}{\phiOne}
        \draw[very thick, ->, blue] (0,0,0) -- (K1) node[anchor=south east]{${k}_1$};
        \draw[dashed, blue!40] (0,0,0) -- (K1xy);
        \draw[dashed, blue!40] (K1) -- (K1xy);

        \tdplotsetcoord{K2}{1}{\thetaTwo}{\phiTwo}
        \draw[very thick, ->, red] (0,0,0) -- (K2) node[anchor=south west]{${k}_2$};
        \draw[dashed, red!40] (0,0,0) -- (K2xy);
        \draw[dashed, red!40] (K2) -- (K2xy);
        
        \tdplotsetthetaplanecoords{\phiOne}
        \tdplotdrawarc[tdplot_rotated_coords, blue, ->]{(0,0,0)}{0.7}{0}{\thetaOne}{anchor=north west}{$\theta_1$}
       
        \tdplotsetthetaplanecoords{\phiTwo}
        \tdplotdrawarc[tdplot_rotated_coords, red, ->]{(0,0,0)}{0.8}{0}{\thetaTwo}{anchor=north west}{$\theta_2$}

        \tdplotsetthetaplanecoords{0} 
        \tdplotdrawarc[dashed, ->]{(0,0,0)}{0.3}{\phiOne}{\phiTwo}{anchor=north}{$\phi$}

    \end{tikzpicture}
    \caption{Illustration of the bispectrum triangle configuration in redshift space, with the LOS aligned along the $z$-axis.}
    \label{fig:1}
\end{figure}

In practice, for a fixed triangle defined by the magnitudes $(k_1, k_2, k_3)$ (or equivalently $k_1, k_2$ and the cosine of the angle between them $\mu_{12}$), the orientation is determined by three angles: the line-of-sight (LOS) angle of the first wavevector $\mu_1 = \cos\theta_1$, the angle between the two wavevectors $\mu_{12} = \hat{k}_1 \cdot \hat{k}_2$, and the azimuthal angle $\phi$ of $\boldsymbol{k}_2$ around $\boldsymbol{k}_1$ (see Figure~\ref{fig:1} as an illustration).
The LOS cosines for the remaining two sides of the triangle are determined by the geometry, which can be calculated by
\begin{equation}
\mu_2 = \mu_1 \mu_{12} + \sqrt{1 - \mu_1^2} \sqrt{1 - \mu_{12}^2} \cos\phi,
\end{equation}
\begin{equation}
\mu_3 = -\frac{k_1 \mu_1 + k_2 \mu_2}{k_3}.
\end{equation}

\section{mock data} \label{sec:mock_data}

\subsection{Simulation}
\label{subsec:simulation}

We utilize the high-resolution, dark-matter-only Jiutian N-body \citep{han2025jiutiansimulationscsstextragalactic,2026SCPMA..6969511Y} simulation suite as the foundation for constructing our mock galaxy catalogs. The specific simulation used in this work has a comoving volume of $1\,(h^{-1}\mathrm{Gpc})^3$ and contains 6144$^3$ particles, yielding a mass resolution of $m_{\rm p}=3.72\times10^8\,h^{-1}\mathrm{M}_\odot$.  The simulation was performed using the L-Gadget3 code, with dark matter halos and their substructures identified via the friends-of-friends and SUBFIND algorithms \citep{SPRINGEL200179,Springel2005}. The fiducial values of the cosmological parameters are set to be the best-fit values from $\it Planck$ results ~\citep{Planck2020}, which are {$h=0.6766$,  $\Omega_{\rm m}=0.3111$,  $\Omega_{\rm b}=0.0490$, $\Omega_{\Lambda}=0.6899$,  $\sigma_{8}=0.8102$,  $n_{\rm s}=0.9665$, $f_{\rm NL}=0$}. The simulation starts at redshift \( z=127 \) and outputs 128 snapshots down to \( z=0 \). This dense temporal sampling allows us to accurately trace the formation and evolution of dark matter structures and provides the flexibility to construct galaxy catalogs at different cosmic epochs.

To account for the effects of RSD and structure evolution, we assemble each simulation cube from multiple slices derived from snapshot outputs at various redshifts. The LOS is aligned parallel to the edge of the simulation box. Slice-like halo catalogs are then combined according to their comoving distances. In our mock catalog, evolutionary effects are addressed by tracing the merger tree of individual galaxies to identify the snapshot with the redshift closest to the galaxy's distance. This method inherently avoids duplicating or omitting galaxies at slice boundaries, in contrast to directly slicing and stitching snapshots based on redshift \citep{Smith_2022a,Smith_2022b}. To maintain the reliability of RSD calculations, we do not apply interpolation when splicing the slices, which has no significant impact on our results at the scales and precision levels of interest. We generate three simulation cubes with central redshifts $z_{\rm c}= [0.3, 0.6, 0.9]$ to construct mock catalogs for galaxies.

{Note that these catalogs retain a plane-parallel cuboid geometry rather than the actual CSST angular footprint and curved-sky light cone. The corresponding survey-window convolution and wide-angle effects are not included in our power-spectrum and bispectrum modeling.}

\subsection{Galaxy mock catalog}
\label{subsec:elg_catalog}
This study focuses on emission line galaxies (ELGs), which constitute one of the primary tracers targeted by the CSST spectroscopic survey. The CSST is equipped with a slitless grism spectrograph operating in three bands ($GU$, $GV$, and $GI$), covering a wavelength range from 225 to 1000 nm. The telescope is planned to observe a sky area of 17,500 deg\textsuperscript{2} over a mission duration of about ten years. For spectroscopic observations, the instrument delivers an angular resolution of approximately 0.3$^{\prime\prime}$, enclosing 80\% of the point-source energy within this diameter. The spectral resolution, defined as $R = \lambda/\Delta\lambda$, exceeds 200. The survey achieves a 5$\sigma$ point-source detection limit of approximately 23 AB magnitude per band.

The mock galaxy catalog is populated using a modified and updated version of the L-Galaxies semi-analytic model \citep{Springel2005,10.1111/j.1365-2966.2005.09675.x,10.1111/j.1365-2966.2006.11287.x,10.1111/j.1365-2966.2010.18114.x}. This version incorporates improvements for modeling the disruption of satellite galaxies and the growth of supermassive black holes compared to the model presented in ~\cite{10.1093/mnras/stv705}. The updated model also introduces new features for characterizing galaxy properties, such as the inclusion of emission line luminosities, which are generated via post-processing techniques ~\citep{Pei2024}. We subsequently use these lines to derive accurate spectroscopic redshifts and to identify galaxies that are detectable by the CSST spectroscopic survey.

For each galaxy, the total redshift $z$ includes contributions from both the cosmological redshift $z_{\text{cos}}$ and the redshift due to peculiar motion $z_{\text{pec}}$, following the relation: $1+z = (1 + z_{\text{cos}})(1 + z_{\text{pec}}) = (1 + z_{\text{cos}})(1 + v_{\text{pec}}/c)$, where $v_{\text{pec}}$ is the line-of-sight component of the galaxy's peculiar velocity. {To account for redshift measurement errors, an independent, zero-mean Gaussian redshift perturbation with standard deviation $\sigma_z = 0.002(1+z)$ is added to each galaxy in the mock catalog \citep[e.g. see][]{Gong_2019}.} {Catastrophic redshift failures and line-confusion interlopers are not considered for simplicity in our analysis.}

We apply a selection criterion to the mock catalog based on the signal-to-noise ratio (SNR) of specific emission lines. The SNR for each galaxy is estimated using four primary optical lines: $\text{H}\alpha$, $\text{H}\beta$, [O\,\textsc{iii}], and [O\,\textsc{ii}]. Given that the spatial extent of an emission line region is typically much smaller than the overall galaxy size, we approximate galaxies as point sources for the purpose of this SNR estimation. For a space-based telescope, the SNR per spectral resolution element for a spectroscopic target can be computed by \citep{10.1093/mnras/sty1980,10.1093/mnras/stac2185}
\begin{equation}
\text{SNR} = \frac{C_{\text{s}} t_{\text{exp}} \sqrt{N_{\text{exp}}}}{\sqrt{C_{\text{s}} t_{\text{exp}} + N_{\text{pix}} \left[ (B_{\text{sky}} + B_{\text{det}}) t_{\text{exp}} + R_{\text{n}}^2 \right]}}.
\end{equation}
The number of detector pixels covered by an object is given by $N_{\text{pix}} = \Delta A / l_p^2$. Here, $\Delta A$ denotes the pixel area on the detector, which, for simplicity, is taken to be the same for all galaxies. The size of the pixels is $l_p = 0.074^{\prime\prime}$, and the point-spread function (PSF) is defined based on the angular resolution of the CSST spectroscopic survey. The number of exposures and the exposure time per visit are set to $N_{\text{exp}} = 4$ and $t_{\text{exp}} = 150$ s, respectively. The read noise and dark current of the detector are $R_n = 5\ e^{-1}\ \text{pixel}^{-1}$ and $B_{\text{det}} = 0.02\ e^{-1}\ \text{pixel}^{-1}$. The sky background, $B_{\text{sky}}$, and the counting rate of a galaxy, $C_s$, are both in the units of $e^{-1}\ \text{pixel}^{-1}$. According to ~\cite{song2024}, the sky background values for the $GU$, $GV$, and $GI$ bands are $B_{\text{sky}} = 0.016$, $0.196$, and $0.266\ e^{-1}\ \text{pixel}^{-1}$, respectively.

We selected galaxies for the mock catalog based on a SNR threshold, requiring that $\text{SNR} \geq 10$ for at least one of the four emission lines (H$\alpha$, H$\beta$, [O\,\textsc{iii}], [O\,\textsc{ii}]) in any of the spectroscopic bands. The resulting galaxy number densities for the three redshift bins centered at $z = 0.3$, $0.6$, and $0.9$ are $\bar{n}_{\rm g} = 1.5 \times 10^{-2}$, $2.1 \times 10^{-3}$, and $4.6 \times 10^{-4} \, h^{3} \text{Mpc}^{-3}$, respectively. These values are in agreement with previous estimates \citep[e.g.][]{Gong_2019}.

\subsection{Edge effect and Random Catalog generation}
We consider the RSD effects and spectroscopic redshift errors in the mock galaxy catalog. This process inevitably causes some galaxies near the boundaries to be shifted out of the simulation volume along the line of sight (LOS), creating artificial ``voids" near the edges. To eliminate these boundary artifacts in the estimation of the power spectrum and the bispectrum, we applied a 50 $h^{-1}\rm Mpc$ radial cut at both edges of the LOS of the mock data. Given that the typical displacement due to redshift-space distortions is about $20-40$ $h^{-1}\rm Mpc$, a 50 $h^{-1}\rm Mpc$ buffer region is sufficiently conservative. Consequently, the volume of the final data catalog is reduced to $1000\times1000\times 900\ (h^{-1}{\rm Mpc})^3$.

For random catalog generation, we employ the ``sample" method rather than the ``shuffle" method. We generate the random catalog by sampling redshifts from a spline interpolation of the comoving number density $n(z)$ measured from the data. Unlike the ``shuffle" method, which preserves the exact observed redshift distribution and can thus lead to a biased normalization (integral constraint), the ``sample" method relies on a smooth and global $n(z)$ derived from spline interpolation. This decouples the random catalog from local density fluctuations, thereby mitigating the impact of the integral constraint.  In modeling $n(z)$, we specifically account for the characteristics of the CSST slitless spectroscopic survey. The selection function $n(z)$ is non-monotonic, exhibiting distinct peaks corresponding to the sensitivity ranges of the $GU$, $GV$, and $GI$ bands and the primary emission lines (e.g. H$\alpha$, H$\beta$, [O \,\textsc{iii}], [O \,\textsc{ii}]).

Furthermore, to minimize the systematic uncertainties arising from the window function convolution, we exclude the fundamental $k$-modes $k_f$ and only consider measurements that start from $k_{\rm min}=2\times k_f$ (more details are discussed in the next subsection). This conservative $k$-cut ensures that the estimated power spectrum and bispectrum are robust against large-scale boundary effects.

\subsection{Power spectrum and Bispectrum estimation}
\label{subsec:measure}

We measure the redshift-space galaxy power spectrum and bispectrum monopoles at three redshifts ($z = 0.3, 0.6, 0.9$) using the \texttt{Triumvirate} pipeline \citep{Wang:2023a,Wang:2023b}.
To maximize the SNR of the clustering measurements, we apply the Feldman-Kaiser-Peacock (FKP) weighting scheme \citep{Feldman_1994}:
\begin{equation}
\label{eq:FKP_weight}
w_{\mathrm{FKP}}(x)=\frac{1}{1+\bar{n}(x)P_{\rm fid}},
\end{equation}
where we adopt a fiducial power spectrum amplitude of $P_{\rm fid}=3\times10^{4}\ (h^{-1}\mathrm{Mpc})^{3}$, corresponding to the characteristic amplitude near the matter-radiation equality scale $k_{\rm eq}$ \citep{Weinberg2008, Dodelson2020}.
Following the weighting step, both the data and random catalogs are assigned to a $128^3$ Cartesian grid using the Triangular Shaped Cloud (TSC) scheme. This discretization mitigates aliasing effects and corresponds to a Nyquist frequency of $k_{N}=0.4\ h\mathrm{Mpc}^{-1}$.
{Before parameter constraints, we subtract the shot noise given by \texttt{Triumvirate} from both $P_0$ and $B_{000}$. The theoretical power spectrum and bispectrum are therefore compared with the corresponding shot-noise-subtracted measurements, and no additional free shot-noise parameters are included in the fitting process. 
}

The choice of scale range is important in the measurement of power spectrum and bispectrum. Since the theoretical framework is based on the tree-level perturbation model, it neglects higher-order loop corrections due to non-linear gravitational instability, non-linear galaxy biasing, and small-scale velocity dispersion effects \citep{Scoccimarro:1997st, Bernardeau:2001qr, Sefusatti:2006eu}. Besides, the validity of the tree-level approximation is redshift-dependent, as non-linearities become more significant at lower redshifts. Hence, we need to restrict the clustering measurements to the linear or quasi-linear regime at different redshifts where the theoretical model remains valid.

\begin{table}[t]
\centering
\caption{Summary of the scale ranges for the galaxy monopole power spectrum $P_0(k)$ and monopole bispectrum $B_{000}(k_1, k_2)$ at different redshifts.}
\label{tab:k_bins}
\begin{tabular}{cccc}
\toprule
Redshift & Statistic & $k_{\rm min}$ [$h\,\rm{Mpc}^{-1}$] & $k_{\rm max}$ [$h\,\mathrm{Mpc}^{-1}$] \\ \midrule
0.3 & $P_0(k)$& 0.014 & 0.119 \\
0.6 & $P_0(k)$& 0.014 & 0.161 \\
0.9 & $P_0(k)$& 0.014 & 0.203 \\ 
All & $B_{000}(k_1, k_2)$ & 0.014 & 0.084 \\ \bottomrule
\end{tabular}
\end{table}

\begin{figure*}
    \centering
    \includegraphics[width=1\linewidth]{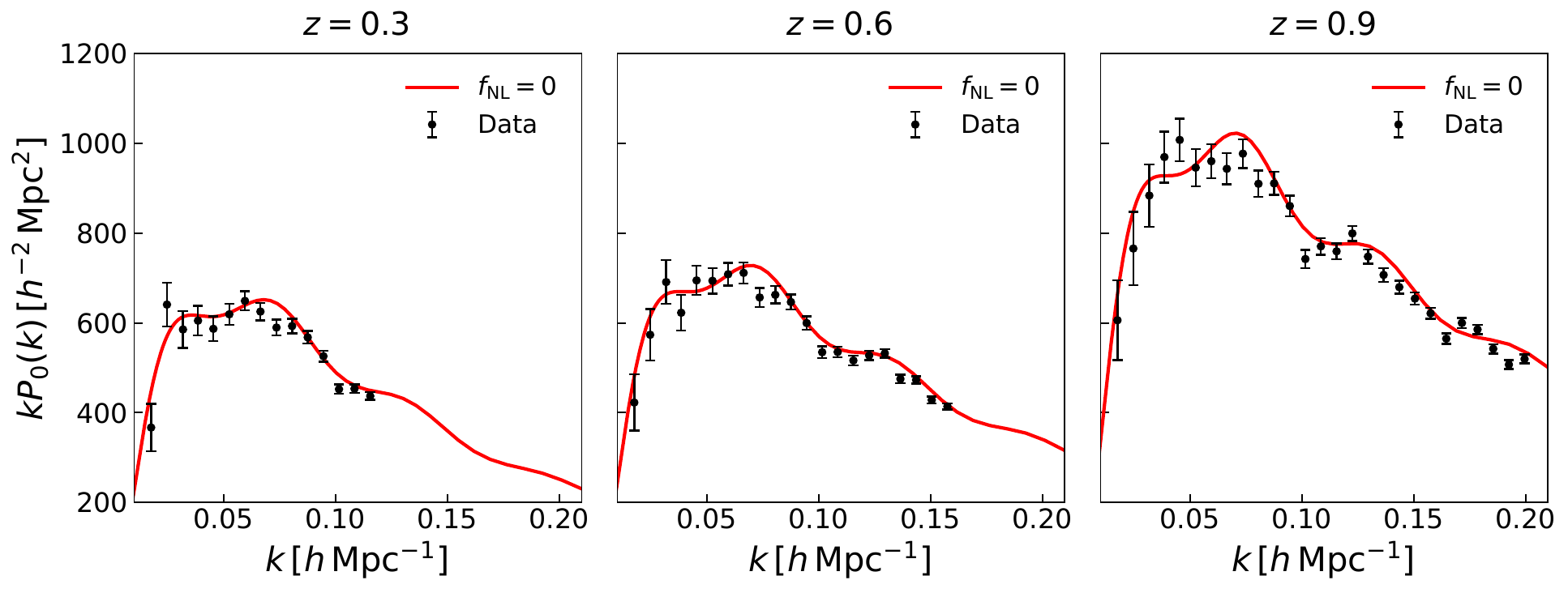}
    \caption{{The monopole power spectrum at $z=0.3$, 0.6, and 0.9. The red curves show the theoretical power spectrum with $f_{\rm NL}=0$. The data points are derived from the simulation, and the error bars are estimated from the analytical covariance matrix. The shot noise has been subtracted at each redshift.}}
    \label{fig:power_measure}
\end{figure*}

\begin{figure}
    \centering
    \includegraphics[width=0.87\linewidth]{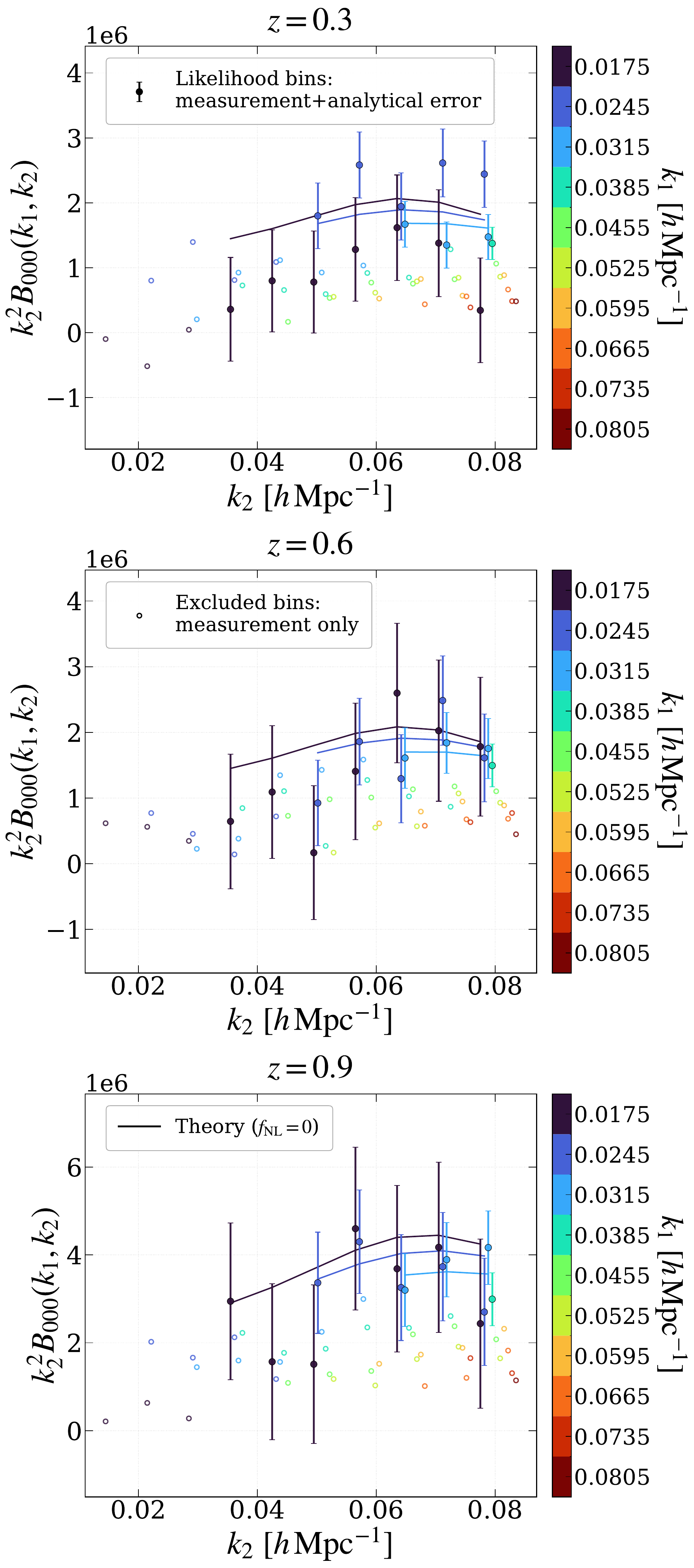}
    \caption{{The monopole bispectrum at $z=0.3$, 0.6, and 0.9. The color bar denotes the value of $k_1$ on the right of each panel. The curves show the theoretical bispectrum with $f_{\rm NL}=0$. The filled data points satisfy $k_2>2k_1$, which are used in the parameter constraints, and the errors bars are estimated by the analytical covariance matrix. The open points indicate the data excluded in the fitting process. The shot noise has been subtracted at each redshift.}}
    \label{fig:bi_measure}
\end{figure}

For the two-point statistics or power spectrum $P(k)$, we adopt redshift-dependent maximum scales $k_{\rm max}$ to align with the tree-level applicability range, based on the characteristic non-linear scales of the Universe at different redshifts \citep{Font-Ribera:2013rwa,Euclid:2019clj}.
For large-scale-cutoff $ k_{\mathrm{min}}$, we use conservative scale cut strategies adopted in standard full-shape analyzes \citep{Gil-Marin:2016,Cabass:2022,Euclid:2026_Linde}, and discard the first fundamental bins $k$ to mitigate large-scale observational systematics and window function artifacts. The minimum scale is set to $ k_{\mathrm{min}} = 2k_f = 0.014 \, h\,\mathrm{Mpc}^{-1} $, where  $k_f = 2\pi/L \approx 0.007 \, h\,\mathrm{Mpc}^{-1}$ is the fundamental frequency for our periodic box. To ensure sufficient sampling of Fourier modes while maintaining computational efficiency, we set a uniform bin width of $\Delta k = 0.007 \, h\,\mathrm{Mpc}^{-1}$ to match $k_f$. The specific choice of the scale ranges for the three redshifts is summarized in Table~\ref{tab:k_bins}.

Regarding the three-point statistics, we measure the monopole of  bispectrum under tripolar spherical harmonic basis: $B_{000}(k_1, k_2)$. Unlike the power spectrum, we apply a more conservative and uniform scale range for the bispectrum across all redshifts to ensure the robust performance of the tree-level bispectrum model, which is typically more sensitive to non-linearities than the power spectrum, and the bispectrum is measured over the range $k \in [0.014, 0.084] \, h\,\mathrm{Mpc}^{-1}$. By restricting the bispectrum to these relatively large scales, we can minimize the impact of loop-level corrections and complex baryonic effects that are not captured by the tree-level template.

Unlike conventional approaches that directly sample discrete triangular configurations in Fourier space \citep{Scoccimarro_2000}, we characterize the redshift-space bispectrum using tri-polar spherical harmonic (TSH) basis decomposition \citep{doi:10.1142/0270}. Implemented via the \texttt{triumvirate} pipeline \citep{Wang:2023a,Wang:2023b}, this formalism expands the bispectrum in terms of TSH rather than relying solely on magnitudes $k_1, k_2, k_3$. The coefficient in the TSH decomposition of the redshift-space bispectrum is given by~\citep{Sugiyama_2018}
\begin{align}
B_{\ell_1\ell_2 L}(k_1,k_2) &= \int\frac{\mathrm{d}\hat{k}_1}{4\pi}\int\frac{\mathrm{d}\hat{k}_2}{4\pi}\int\frac{\mathrm{d}\hat{n}}{4\pi} \mathcal{W}_{\ell_1\ell_2 L}(\hat{k}_1,\hat{k}_2,\hat{n}) \nonumber \\
&\quad \times B(\boldsymbol{k}_1,\boldsymbol{k}_2,-\boldsymbol{k}),
\label{eq:tri_polar_coefficient}
\end{align}

\begin{align}
\label{eq:Wigner_3j_weight}
\mathcal{W}_{\ell_1\ell_2 L}(\hat{k}_1,\hat{k}_2,\hat{n}) &\equiv (2\ell_1+1)(2\ell_2+1)(2L+1) \nonumber \\
&\quad {\times}
\begin{pmatrix}
\ell_1 & \ell_2 & L \\
0 & 0 & 0
\end{pmatrix} \nonumber \\
&\quad \times \sum_{m_1,m_2,M} 
\begin{pmatrix}
\ell_1 & \ell_2 & L \\
m_1 & m_2 & M
\end{pmatrix} \nonumber \\
&\quad \times y_{\ell_1}^{m_1}(\hat{k}_1) y_{\ell_2}^{m_2}(\hat{k}_2) y_L^M(\hat{n}),
\end{align}
where $\boldsymbol{k} = \boldsymbol{k}_1 + \boldsymbol{k}_2$, $y_{\ell}^{m} = \sqrt{4\pi/(2\ell+1)} \, Y_{\ell}^{m}$ is a normalized spherical harmonic function, and the matrix with six indices represents the Wigner-3j symbol. In this decomposition, the index $L$ governs the expansion with respect to the LOS direction. The mode of $B_{\ell_1\ell_2 L}$ with $L=0$ describes isotropic components in the bispectrum, whereas the modes with $L>0$ arise from anisotropic components alone. Hence, we refer to $B_{\ell_1\ell_2 L}$ with $L=0$ and $L=2$ as the monopole and quadrupole bispectra, respectively.

Specifically, we measure the monopole bispectrum $B_{000}(k_1, k_2)$, which represents the leading-order term in the TSH expansion. This statistic is obtained by integrating and averaging both the orientation of the triangle relative to the line-of-sight and the internal angle between the wavevectors $\boldsymbol{k}_1$ and $\boldsymbol{k}_2$. Consequently, the measured $B_{000}$ incorporates information from all possible triangular configurations that satisfy the triangle inequality, $|k_1 - k_2| \le k_3 \le k_1 + k_2$.
The monopole bispectrum is then computed via a integration over the angular grid, which is given by
\begin{equation}
\begin{aligned}
B_{000}(k_1, k_2) = &\frac{1}{8\pi} \int_{-1}^1 d\mu_{12} \int_{-1}^1 d\mu_1 \int_0^{2\pi} d\phi \, \\
&\times B_g(k_1, k_2, \mu_{12}, \mu_1, \phi),
\end{aligned}
\end{equation}
where the integration accounts for the tree-level kernels $Z_1, Z_2$ and the FoG suppression evaluated in each angular configuration as shown in Equation~(\ref{eq:Bg}).

Our primary focus is to constrain the local-type PNG parameter, $f_{\mathrm{NL}}$. Physically, the signal for local-type PNG is maximally enhanced in the ``squeezed" limit, where one wavelength is significantly larger than the others (i.e. $k_1 \ll k_2 \approx k_3$). To ensure that our parameter estimation is dominated by the configurations most sensitive to local-type PNG and to maintain the validity of our tree-level model, we implement a specific configuration cut. During the likelihood analysis, we restrict our data points to those satisfying the condition $k_2 > 2k_1$. By imposing this scale separation, we effectively isolate the squeezed-limit information necessary for a robust probe of local-type PNG while mitigating the impact of configurations where other types of non-Gaussianity or non-linear gravitational effects might dominate.

{In Figure~\ref{fig:power_measure} and Figure~\ref{fig:bi_measure}, we show the data points of the  monopole power spectra and bispectra measured from the mock CSST galaxy catalog at the three redshifts. The error bars are obtained from the analytical covariance matrix described in the next section.}
{In Figure~\ref{fig:bi_measure}, we notice that some bispectrum data points lie below the theoretical curves, although most of them agree with the curves within $1\sigma$ uncertainties. These offsets are mainly due to finite-volume statistical fluctuations in the simulation. The galaxy sample at each redshift has a volume of $\sim 0.9\,(h^{-1}\mathrm{Gpc})^3$, and the constraint results reported in this work are derived from this volume rather than the full CSST survey volume.}

\subsection{Covariance Matrix Estimation}
\label{sec:covariance}

\begingroup
\begin{figure*}
    \centering
    \includegraphics[width=1\linewidth]{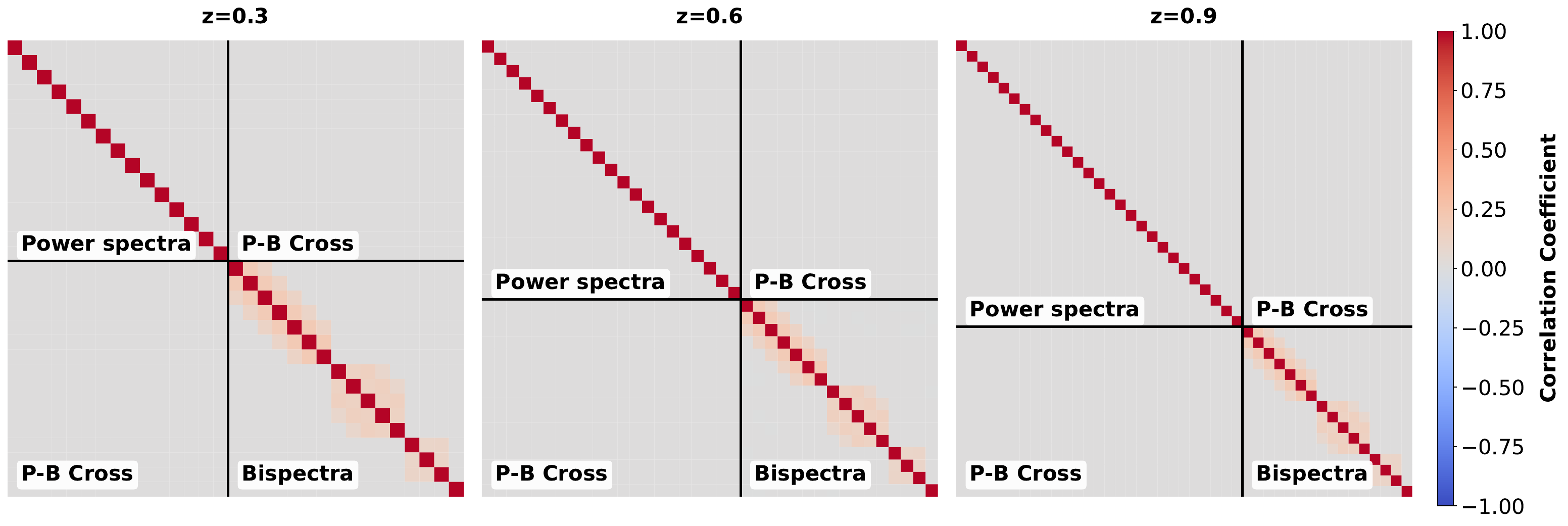}
    \caption{Normalized joint theoretical covariance matrices of the power spectrum ($P_0$) and bispectrum ($B_{000}$) at $z=0.3$ (left), $0.6$ (middle), and $0.9$ (right). The matrices are computed analytically using tree-level standard perturbation theory, comprising 15 $P$ bins + 16 $B$ bins at $z=0.3$, 21 $P$ + 16 $B$ bins at $z=0.6$, and 27 $P$ + 16 $B$ bins at $z=0.9$, explicitly targeting the squeezed-limit configurations.}
    \label{fig:cov}
\end{figure*}

To robustly evaluate the likelihood for $f_{\rm NL}$, a precise estimation of the covariance matrix is required, which needs to account for the correlations between the power spectrum and the bispectrum, as well as the cross-correlations across different scale bins. Instead of relying on empirical sample covariances from limited mock catalogs, which suffer from statistical noise and necessitate the Hartlap correction factor when inverting the precision matrix, we analytically construct the full theoretical covariance matrix based on Standard Perturbation Theory (SPT) \citep{Sugiyama2020MNRAS}.

The joint data vector at each redshift concatenates the monopole power spectrum and bispectrum: $\mathbf{D} = [P_0(k_i), B_{000}(k_j, k_k)]^T$. The grand covariance matrix $\mathbf{C}$ is strictly defined as the ensemble average of the data vector fluctuations:
\begin{equation}
    \mathbf{C} \equiv \langle \mathbf{D}\mathbf{D}^T \rangle - \langle \mathbf{D} \rangle \langle \mathbf{D} \rangle^T = 
    \begin{pmatrix} 
    \mathbf{C}^{PP} & \mathbf{C}^{PB} \\ 
    \mathbf{C}^{BP} & \mathbf{C}^{BB} 
    \end{pmatrix}.
\end{equation}
This block matrix naturally decomposes into the power spectrum auto-covariance ($\mathbf{C}^{PP}$), the bispectrum auto-covariance ($\mathbf{C}^{BB}$), and the cross-covariance ($\mathbf{C}^{PB} = (\mathbf{C}^{BP})^T$).

The analytical computation of these covariance blocks requires theoretical templates for the underlying anisotropic power spectrum $P_g(\mathbf{k})$ and bispectrum $B_g(\mathbf{k}_1, \mathbf{k}_2, \mathbf{k}_3)$. We evaluate these using the exact linear matter power spectrum $P_{\rm lin}(k)$ generated by the \texttt{CLASS} code at the fiducial cosmology. 
By applying Wick's theorem and the perturbative expansion to the definitions above, we formulate the integral representations of each covariance block. 

\paragraph{Power Spectrum Auto-covariance:}
On quasi-linear scales, the power spectrum covariance $\mathbf{C}^{PP}$ is heavily dominated by its Gaussian disconnected component. The variance for a given bin is purely diagonal and explicitly depends on the number of available independent Fourier modes $N_k$, which is given by
\begin{equation}
    \mathbf{C}^{PP}(k, k') \simeq \mathbf{C}^{PP}_{\rm Gauss}(k, k') = \frac{2}{N_{k}} \left[ P_0(k) + \frac{1}{\bar{n}} \right]^2 \delta_{k, k'}.
\end{equation}

\paragraph{Bispectrum Auto-covariance:}
The bispectrum covariance $\mathbf{C}^{BB}$ is fundamentally driven by the 6-point function of the density field, which expands hierarchically into
\begin{equation}
    \mathbf{C}^{BB} = \mathbf{C}^{BB}_{PPP} + \mathbf{C}^{BB}_{BB} + \mathbf{C}^{BB}_{PT} + \mathbf{C}^{BB}_{P_6}.
\end{equation}
The dominant contribution arises from the zero-leg shared Gaussian disconnected term, $\mathbf{C}^{BB}_{PPP} \propto P_g(\mathbf{k}_1) P_g(\mathbf{k}_2) P_g(\mathbf{k}_3) / V$, which governs the vast majority of the signal (see the diagonals in Figure~\ref{fig:cov}). To capture the non-Gaussian mode coupling, we include the leading single-leg shared term $\mathbf{C}^{BB}_{BB}$, which scales as $\propto B_g(\mathbf{k}_3, \mathbf{k}_1, \mathbf{k}_2) B_g(\mathbf{k}'_3, \mathbf{k}'_1, \mathbf{k}'_2) / V$ and predominantly sources the off-diagonal structure. By explicitly evaluating the leading higher-order corrections (such as the two-leg shared trispectrum $\mathbf{C}^{BB}_{PT}$ involving the $Z_3$ and $Z_4$ kernels), we verified that the remaining connected terms, including the fully connected six-point function $\mathbf{C}^{BB}_{P_6}$ (which would formally require up to the $Z_5$ kernel), are numerically subdominant for the squeezed-limit configurations considered here across all redshift bins. We therefore retain the $PPP$ and leading single-leg shared $BB$ terms in the bispectrum covariance used in this analysis.

\paragraph{Cross-covariance:}
The cross-covariance $\mathbf{C}^{PB}$ directly links the power spectrum and the bispectrum through the 5-point function. Neglecting the cross-covariance can misestimate the joint parameter uncertainties and alter the resulting posterior. Following the terminology in \cite{Sugiyama2020MNRAS}, we split this cross-covariance into a disconnected $P \times B$ term and a connected five-point function ($P_5$) term. The leading contribution is driven by the disconnected $P \times B$ component via a single shared momentum leg, which scales as
\begin{equation}
\begin{aligned}
    \mathbf{C}^{PB}(k_1; k'_1, k'_2, k'_3) &\simeq \mathbf{C}^{PB}_{PB} \\
    &\propto \frac{1}{V} B_g(\mathbf{k}'_1, \mathbf{k}'_2, \mathbf{k}'_3) P_g(\mathbf{k}_1).
\end{aligned}
\end{equation}
All the above covariance blocks are evaluated via a highly optimized vectorized Monte Carlo engine. Millions of triangle configurations are sampled uniformly within the exact bin geometries---specifically targeting the $k_2 > 2k_1$ squeezed-limit filter crucial for $f_{\rm NL}$ constraints. Importantly, because the covariance is strictly evaluated at the fiducial model where $f_{\rm NL} = 0$, the non-Gaussian scale-dependent bias vanishes (i.e., $\Delta b(k) = 0$). Consequently, the covariance matrix remains completely independent of the tracer-response parameter $p$ introduced in Eq.~(12), decoupling the theoretical covariance generation from the variations of the likelihood parameter space.

\endgroup

\section{Constraints and Results}
\label{sec:results}

To constrain the PNG parameter $f_{\rm NL}$ with other free parameters, we perform a joint analysis of the galaxy power spectrum and bispectrum across three distinct redshift bins.
For each redshift $z_i$, we define a joint data vector $\mathbf{D}_i$, which concatenates the measured monopole power spectrum and bispectrum:
\begin{equation}
\mathbf{D}_i = [P_0(k), B_{000}(k_1, k_2)]_{z_i}.
\end{equation}
The log-likelihood for a given redshift is given by
\begin{equation}
\ln \mathcal{L}_i (\theta) \propto -\frac{1}{2} [\mathbf{D}_i - \mathbf{T}_i(\theta)]^T \mathbf{C}_i^{-1} [\mathbf{D}_i - \mathbf{T}_i(\theta)],
\end{equation}
where $\mathbf{T}_i(\theta)$ is the theoretical model evaluated given the parameters $\theta$, and $\mathbf{C}_i$ is the joint covariance matrix for the $i$-th redshift, which accounts for the cross-covariance between the power spectrum and the bispectrum at the same redshift, ensuring that the information is not double-counted and that the correlations between different scales and probes are properly captured. The total log-likelihood for all samples at different redshifts is the sum of the individual log-likelihoods
\begin{equation}
\ln \mathcal{L}_{\text{tot}}(\theta) = \sum_{i=1}^{3} \ln \mathcal{L}_i(\theta).
\end{equation}
We then explore the multi-dimensional parameter space using the Markov Chain Monte Carlo (MCMC) method to obtain the posterior distributions of the model parameters.

\begin{table*}[t]
\centering
\caption{{Priors, best-fit values, and 1$\sigma$ errors of the model parameters for the $P$-only and $P+B$ analyses in the $p=0.55$ and $p=1$ cases.}}
\label{tab:combined_results}
\begin{tabular}{llcccc}
\hline \hline
Parameter & Prior & \multicolumn{2}{c}{{$p=0.55$}} & \multicolumn{2}{c}{{$p=1$}} \\
\cmidrule(lr){3-4}\cmidrule(lr){5-6}
 & & $P$-only & $P+B$ & {$P$}-{only} & {$P+B$} \\
\hline
\multicolumn{6}{c}{\textbf{{Global parameter}}} \\
$f_{\rm NL}$ & $\mathcal{U}[-500,500]$ & {$-4\pm43$} & {$3\pm41$} & {$-3\pm86$} & {$54^{+70}_{-80}$} \\
\hline
\multicolumn{6}{c}{\textbf{{Redshift bin $z=0.3$}}} \\
$b_1$ & $\mathcal{U}[0,10]$ & {$0.940\pm0.015$} & {$0.941\pm0.015$} & {$0.939\pm0.013$} & {$0.943\pm0.012$} \\
$b_2$ & $\mathcal{U}[-10,10]$ & $-$ & {$-2.53^{+0.14}_{-0.19}$} & {$-$} & {$-2.45^{+0.16}_{-0.22}$} \\
$\sigma_P$ & $\mathcal{U}[0,10]$ & {$6.21\pm0.56$} & {$6.10\pm0.56$} & {$6.19\pm0.53$} & {$6.17\pm0.53$} \\
$\sigma_B$ & $\mathcal{U}[0,10]$ & $-$ & {$3.7^{+1.7}_{-3.0}$} & {$-$} & {$3.8^{+1.9}_{-3.1}$} \\
\hline
\multicolumn{6}{c}{\textbf{{Redshift bin $z=0.6$}}} \\
$b_1$ & $\mathcal{U}[0,10]$ & {$1.147\pm0.015$} & {$1.149\pm0.014$} & {$1.146^{+0.011}_{-0.012}$} & {$1.147\pm0.011$} \\
$b_2$ & $\mathcal{U}[-10,10]$ & $-$ & {$-3.14^{+0.19}_{-0.23}$} & {$-$} & {$-3.01\pm0.27$} \\
$\sigma_P$ & $\mathcal{U}[0,10]$ & {$4.14\pm0.33$} & {$4.09\pm0.33$} & {$4.13\pm0.30$} & {$4.05\pm0.30$} \\
$\sigma_B$ & $\mathcal{U}[0,10]$ & $-$ & {$3.3^{+1.5}_{-2.8}$} & {$-$} & {$3.3^{+1.5}_{-2.8}$} \\
\hline
\multicolumn{6}{c}{\textbf{{Redshift bin $z=0.9$}}} \\
$b_1$ & $\mathcal{U}[0,10]$ & {$1.628\pm0.021$} & {$1.631\pm0.021$} & {$1.628\pm0.024$} & {$1.619\pm0.020$} \\
$b_2$ & $\mathcal{U}[-10,10]$ & $-$ & {$-3.31^{+0.30}_{-0.47}$} & {$-$} & {$-3.17^{+0.36}_{-0.45}$} \\
$\sigma_P$ & $\mathcal{U}[0,10]$ & {$2.80^{+0.30}_{-0.27}$} & {$2.75^{+0.30}_{-0.27}$} & {$2.80^{+0.32}_{-0.28}$} & {$2.63^{+0.31}_{-0.27}$} \\
$\sigma_B$ & $\mathcal{U}[0,10]$ & $-$ & {$4.2^{+2.0}_{-3.6}$} & {$-$} & {$4.2^{+2.0}_{-3.4}$} \\
\hline
\end{tabular}
\end{table*}

{For power spectrum only ($P$-only) constraints, we set three free parameters per redshift as ($f_{\mathrm{NL}}$, $b_1$, $\sigma_{P}$), and for combined power spectrum and bispectrum ($P+B$) constraints, we set five free parameters per redshift as ($f_{\mathrm{NL}}$, $b_1$, $b_{2}$, $\sigma_{P}$, $\sigma_{\text{B}}$).} {Here, $f_{\rm NL}$ is shared across redshifts, while the other parameters vary independently, giving seven and 13 parameters in the joint-redshift analyses, respectively.} We adopt flat priors for all the free parameters, which are listed in Table~\ref{tab:combined_results}.  All other cosmological parameters, i.e., {$h$,  $\Omega_{\rm m}$,  $\Omega_{\rm b}$, $\Omega_{\Lambda}$, $\sigma_{8}$, and $n_{\rm s}$}, are fixed to the $\it Planck$ best-fit values \citep{Planck2020}.

The MCMC process is implemented using the affine-invariant ensemble sampler \texttt{emcee} and its differential-evolution proposal \texttt{DEMove} \citep{Foreman_Mackey_2013}. Convergence is assessed using the integrated autocorrelation time criterion, ensuring that the chain length exceeds 50 times the autocorrelation time for all parameters.

\begin{figure}
    \centering
    \includegraphics[width=1\linewidth]{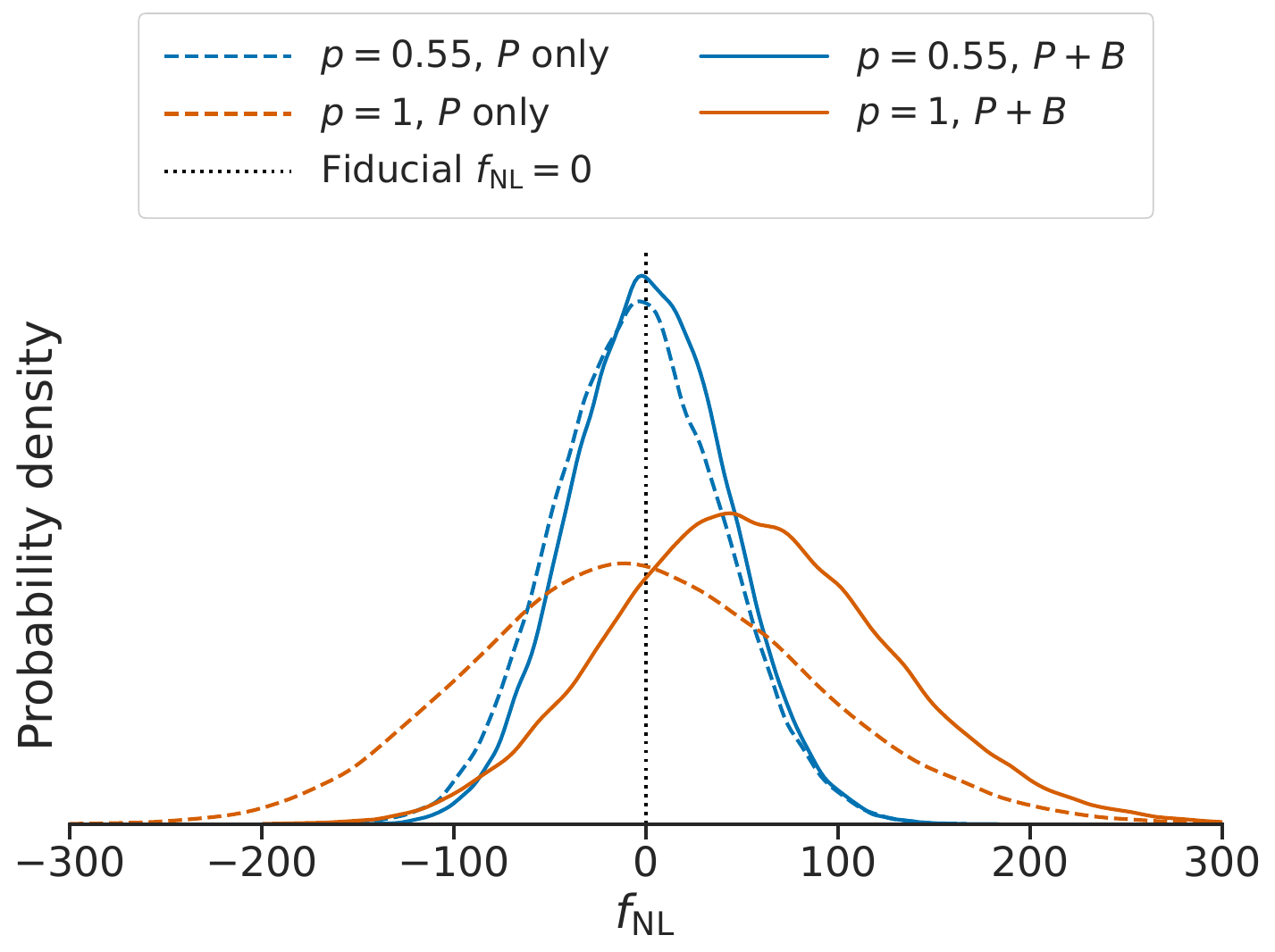}
    \caption{{Marginalized one-dimensional posteriors of $f_{\rm NL}$. Dashed and solid curves show the results from $P$-only and $P+B$ analyses, respectively. Blue and orange curves denote the $p=0.55$ and $p=1$ cases, respectively, and the vertical dotted line marks the fiducial value $f_{\rm NL}=0$.}}
    \label{fig:p+b}
\end{figure}

In Figure~\ref{fig:p+b}, we show the constraint results of the one-dimensional (1-d) probability distribution functions (PDFs) for the PNG parameter $f_{\rm NL}$ using the mock data of $P$-only and $P$+$B$. {For the  $p=0.55$ case, we obtain $f_{\rm NL}=-4\pm43$ and $3\pm41$ at 68\% confidence level (CL) for the $P$-only and $P$+$B$ cases, respectively, as shown in Table~\ref{tab:combined_results}.} {We can find that the constraints on $f_{\rm NL}$ are consistent with the fiducial value $f_{\rm NL}=0$ within 1$\sigma$, with an improvement of 4.7\% in the constraint precision for the $P$+$B$ case compared to the $P$-only case.
The modest gain is mainly due to the usable and conservative scales of the bispectrum data in our analysis, which are limited by the simulation size. For a real Stage~IV survey with large sky coverage and high-accuracy large-scale data, the bispectrum will play a more important role in the PNG constraint, since its main constraint power comes from large scales.
For $p=1$, we obtain $f_{\rm NL}=-3\pm86$ from the $P$-only case and $54^{+70}_{-80}$ from the $P$+$B$ joint analysis. The weaker constraints arise because $b_\phi=2\delta_c(b_1-p)$ is suppressed more strongly in this case compared to $p=0.55$. For the biased best-fit value of $f_{\rm NL}$ in the $P$+$B$ with $p=1$ case, after performing the tests with 200 data realizations {(the details are described in Appendix~\ref{app:mock_recovery})}, we confirm that it is mainly due to statistical fluctuations of the data and parameter degeneracies with the nuisance parameters. The current result in this case is reasonable, and we may not achieve better agreement in the current simulation analysis.

{Note that these results are obtained using the mock data within a $0.9\,(h^{-1}{\rm Gpc})^3$ retained analysis volume at each redshift. For the full CSST spectroscopic survey covering $17,500~\mathrm{deg}^{2}$, scaling the joint constraints from the sum of these three analysis volumes to $31.7\,(h^{-1}{\rm Gpc})^3$ over $0.1<z<1.2$ with $\sigma\propto V^{-1/2}$ gives an indicative $\sigma(f_{\rm NL})\simeq12$--22 \citep{Gong_2019,miao2024forecastingbaomeasurementscsst}.}
The constraint results of other parameters are listed in Table~\ref{tab:combined_results}, and the 1-d PDFs and contour maps of all free parameters are shown in the Appendix.

It is important to note that the sensitivity to $f_{\rm NL}$ is inherently related to the tracer's response to primordial potential fluctuations, characterized by the parameter $b_{\phi}=2\delta_{c}(b_{1}-p)$.  For ELGs, i.e. the primary tracer in this work, the linear bias $b_{1}$ is typically lower than that of Luminous Red Galaxies (LRGs) or Quasars (QSOs).  This leads to a smaller $b_{\phi}$ and a correspondingly suppressed scale-dependent bias signal. Consequently, ELGs are relatively less effective for constraining $f_{\rm NL}$ compared to high-bias populations.

To overcome this limitation and further enhance the constraints on $f_{\rm NL}$, the inclusion of multiple types of cosmic tracers offers a promising avenue.  The multi-tracer technique~\citep{PhysRevLett.102.021302} allows for the effective cancellation of cosmic variance by cross-correlating different tracers that share the same underlying dark matter density field.  Since the CSST spectroscopic survey is expected to observe a diverse population of galaxies including ELGs, LRGs, and high-redshift QSOs \citep{miao2024forecastingbaomeasurementscsst}, performing a joint multi-tracer analysis would allow us to measure the ratio of their clustering amplitudes independent of stochastic density fluctuations.  This synergy could lead to a substantial improvement in the precision of $f_{\rm NL}$, potentially surpassing the statistical limits of any single-tracer analysis. We will study these effects in our future work.

\section{CONCLUSION} \label{sec:conclusion}

In this study, we have developed a comprehensive analysis framework for the redshift-space galaxy power spectrum and bispectrum to study the constraints on the PNG in the upcoming CSST spectroscopic survey. Using the high-resolution Jiutian N-body simulation suite, we construct CSST-like mock catalogs for ELGs at $z=0.3$, 0.6, and 0.9, which explicitly incorporate selected CSST observational characteristics.

We consider different effects in the theoretical modeling, including the linear Kaiser effect, FoG damping, and scale-dependent bias induced by $f_{\rm NL}$. {For the three-point statistics, we employ the tree-level perturbation theory and utilize the tripolar spherical harmonic decomposition to measure and model the bispectrum monopole $B_{000}$.} {Through a joint MCMC analysis using an analytical covariance, we demonstrate that while the power spectrum alone can provide competitive bounds, the inclusion of the bispectrum, specifically targeting squeezed-enriched configurations, improves the $f_{\rm NL}$ constraint precision by approximately 4.7\% for $p=0.55$.} {Our joint analysis yields $f_{\rm NL}=3\pm41$ for $p=0.55$ and $54^{+70}_{-80}$ for $p=1$, and the constraint on $f_{\rm NL}$ for the full CSST survey is expected to reach $\sigma(f_{\rm NL})\simeq12$--22.}

In conclusion, the CSST spectroscopic survey is expected to provide constraints on the PNG, and thereby help test inflation models. Especially, when cooperating with other surveys, such as $\it Euclid$, DESI, and the Roman Space Telescope (RST) \citep{spergel2015widefieldinfrarredsurveytelescopeastrophysics,Eifler_2021}, the synergy may help in mitigating systematic uncertainties and improving constraints on the initial conditions and physics of the early Universe.

\begin{acknowledgments}
J.N.D. and Y.G. acknowledge the support from the CAS Project for Young Scientists in Basic Research (No. YSBR-092), and National Key R\&D Program of China grant Nos. 2022YFF0503404 and 2020SKA0110402. X.L.C. acknowledges the support of the National Natural Science Foundation of China through grant Nos. 11473044 and 11973047 and the Chinese Academy of Science grants ZDKYYQ20200008, QYZDJ- SSW-SLH017, XDB 23040100, and XDA15020200. Q.G. acknowledges the support from the National Natural Science Foundation of China (NSFC No. 12033008). H.T.M. acknowledges the support from the National Natural Science Foundation of China (NSFC, grant No. 12503008). The Jiutian simulations were conducted under the support of the science research grants from the China Manned Space Project with grant No. CMS- CSST-2021-A03. This work is also supported by science research grants from the China Manned Space Project with grant Nos. CMS-CSST-2025-A02, CMS-CSST-2021-B01, and CMS-CSST-2021-A01.
\end{acknowledgments}

\appendix
\section{Full posterior distributions}
\label{append}
In Figures~\ref{fig:3reds} and \ref{fig:p+b contour}, we show the full posterior distributions for $p=0.55$ and $p=1$ in the power-spectrum-only and joint power-spectrum--bispectrum analyses, respectively.

\begin{figure*}
    \centering
    \includegraphics[width=1\linewidth]{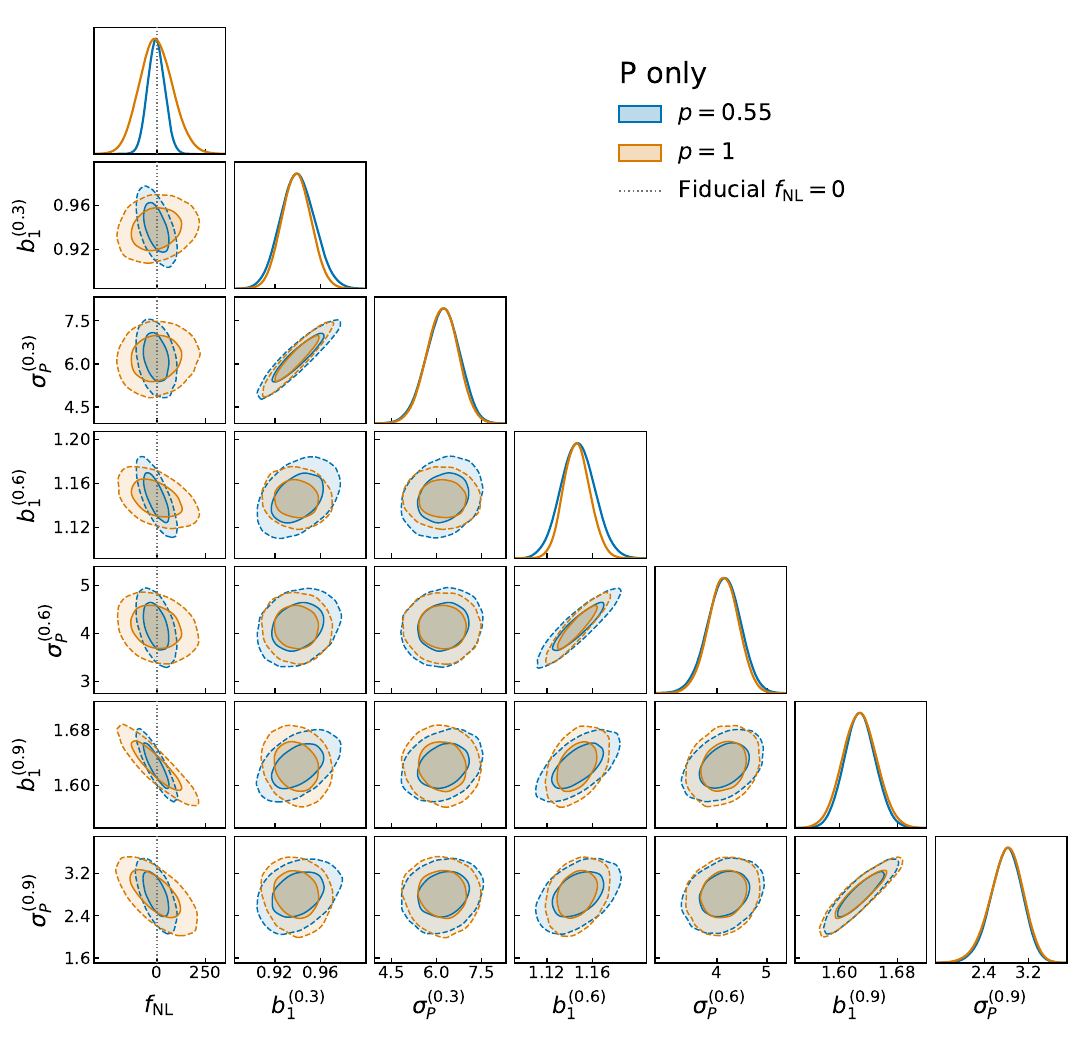}
    \caption{{Marginalized one- and two-dimensional posterior densities for the power-spectrum-only analysis of the three redshift samples, using the analytical covariance. Blue and orange denote $p=0.55$ and $p=1$, respectively. Solid and dashed contours enclose 68\% and 95\% CLs of the smoothed two-dimensional densities; diagonal panels show the one-dimensional densities. The dotted lines mark the simulation input $f_{\rm NL}=0$. Superscripts on the nuisance parameters indicate the redshift.}}
    \label{fig:3reds}
\end{figure*}

\begin{figure*}
    \centering
    \includegraphics[width=1\linewidth]{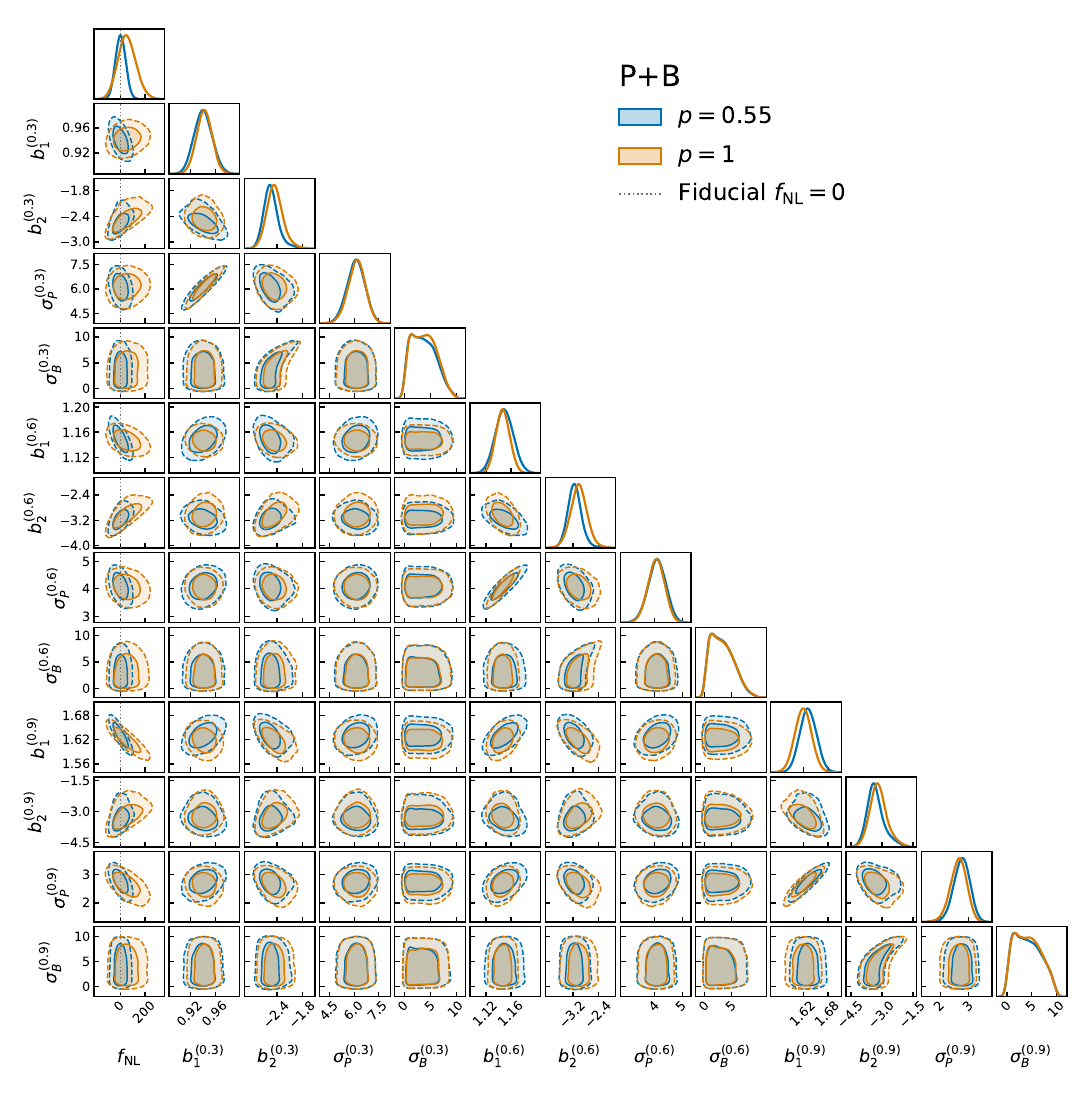}
    \caption{{Marginalized one- and two-dimensional posterior densities for all 13 parameters of the joint power-spectrum--bispectrum analysis of the three redshift samples, using the analytical covariance. Colors, contour probabilities, and line styles follow Figure~\ref{fig:3reds}. The parameter $f_{\rm NL}$ is shared across redshifts, while $b_1$, $b_2$, $\sigma_P$, and $\sigma_B$ are independent in each redshift bin. }
}
    \label{fig:p+b contour}
\end{figure*}
\clearpage

\section{Recovery tests with 200 data realizations}
\label{app:mock_recovery}

{We use recovery tests to examine how statistical fluctuations and nuisance
parameters affect the $f_{\rm NL}$ constraints as shown in Figure~\ref{fig:p+b}. We first jointly
fit the power spectrum and bispectrum measured from the simulation to constrain the nuisance parameters, i.e. $b_1$, $b_2$, $\sigma_P$, and $\sigma_B$, by fixing $f_{\rm NL}=0$. We then evaluate the theoretical model using these best-fit values to obtain a ``noiseless'' data vector, $\boldsymbol d_0$. This construction provides a known input
$f_{\rm NL}=0$ and a specified set of nuisance inputs for the recovery tests. Starting from $\boldsymbol d_0$, we generate 200 data vectors as 200 realizations, $\boldsymbol d_i=\boldsymbol d_0+\boldsymbol\epsilon_i$, by drawing independent Gaussian noise vectors $\boldsymbol\epsilon_i\sim\mathcal N(\boldsymbol 0,\boldsymbol C)$. Here $\boldsymbol C$ is the theoretical covariance matrix as discussed in Section~3.5. Each data vector is then fitted for the $P$-only and $P+B$ cases with $p=1$ and $0.55$, and $f_{\rm NL}$ and the nuisance parameters are set as free parameters in the fitting process.}

{In Figure~\ref{fig:mock_paired}, we show the best-fit values of $f_{\rm NL}$ for the 200 data realizations (blue dots) from the $P$-only and $P+B$ cases with $p=1$ and $p=0.55$. We find that the distribution of the best-fit $f_{\rm NL}$ for the $p=1$ case is much more dispersed than that for the $p=0.55$ case, which is consistent with the results shown in Figure~\ref{fig:p+b}. We also calculate the mean of the 200 data vectors and obtain a mean data vector. By fitting the mean data vector with the covariance $\boldsymbol C$, we find the best-fit $f_{\rm NL}=9.46$ and 9.95 for $P$-only and $P+B$ with $p=1$, and 3.38 and 3.76 for $p=0.55$ (open circles), which are very close to the fiducial value $f_{\rm NL}=0$. The error bars of the open circles are derived as the standard deviations of the best-fit $f_{\rm NL}$ of the 200 realizations, and we can find that the best-fit values of $f_{\rm NL}$ from the simulation (purple star) are within the error bars. These tests indicate that the biased best-fit values of $f_{\rm NL}$ shown in Figure~\ref{fig:p+b} are mainly due to statistical fluctuations of the data and parameter degeneracies, which is statistically acceptable in the analysis.}

\begin{figure*}
	\centering
	\includegraphics[width=\linewidth]{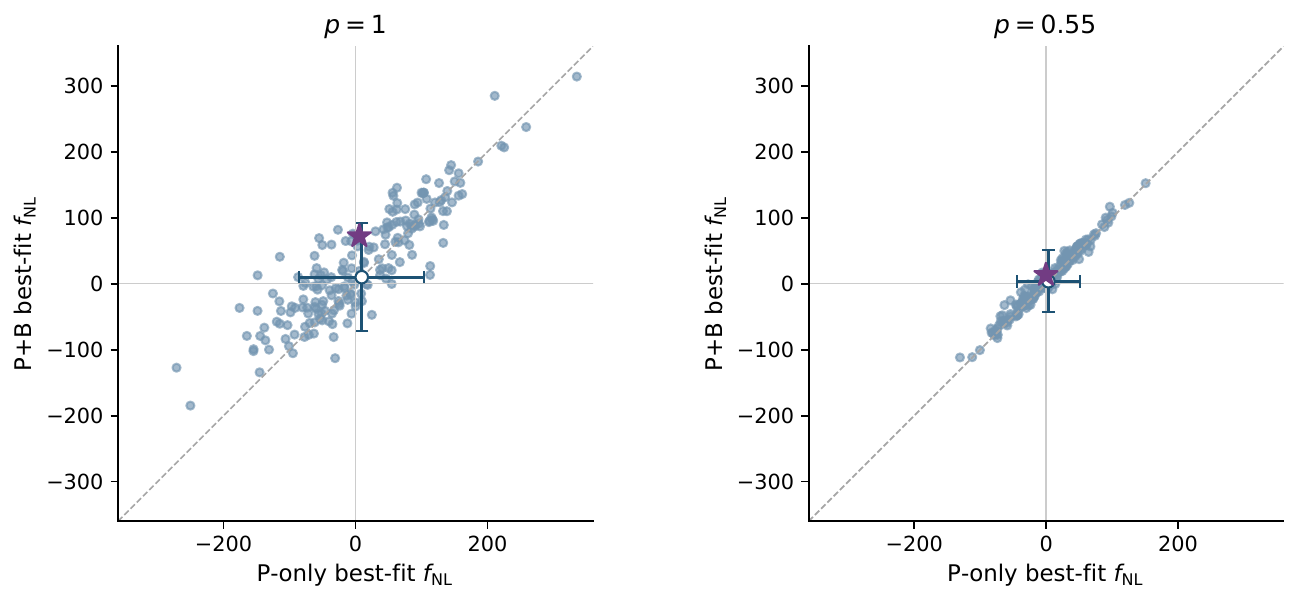}
	\caption{The best-fit values of $f_{\rm NL}$ from the $P$-only and $P+B$ cases with $p=1$ (left) and $p=0.55$ (right) for the 200 realizations. Each blue point represents the best-fit $f_{\rm NL}$ of one realization. The open circle shows the best-fit $f_{\rm NL}$ using the mean data vector of the  200 realizations with the theoretical covariance matrix. Its error bars give the standard deviations of the 200 individual best fits along each axis. The purple star shows the best-fit $f_{\rm NL}$ derived from  the simulation. The dashed diagonal marks equal $P$-only and $P+B$ estimates, and the grey solid lines indicate the fiducial value $f_{\rm NL}=0$. }
	\label{fig:mock_paired}
\end{figure*}

\bibliography{sample701}{}

\begin{thebibliography}{}
\expandafter\ifx\csname natexlab\endcsname\relax\def\natexlab#1{#1}\fi
\providecommand{\url}[1]{\href{#1}{#1}}
\providecommand{\dodoi}[1]{doi:~\href{http://doi.org/#1}{\nolinkurl{#1}}}
\providecommand{\doeprint}[1]{\href{http://ascl.net/#1}{\nolinkurl{http://ascl.net/#1}}}
\providecommand{\doarXiv}[1]{\href{https://arxiv.org/abs/#1}{\nolinkurl{https://arxiv.org/abs/#1}}}

\bibitem[{M. Alvarez {et~al.}(2014)Alvarez, Baldauf, Bond, Dalal, de~Putter,
  Doré, Green, Hirata, Huang, Huterer, Jeong, Johnson, Krause, Loverde,
  Meyers, Meerburg, Senatore, Shandera, Silverstein, Slosar, Smith,
  Zaldarriaga, Assassi, Braden, Hajian, Kobayashi, Stein, \& van
  Engelen}]{alvarez2014testinginflationlargescale}
Alvarez, M., Baldauf, T., Bond, J.~R., {et~al.} 2014, Testing Inflation with
  Large Scale Structure: Connecting Hopes with Reality, \doarXiv{1412.4671}

\bibitem[{T. Baldauf {et~al.}(2011)Baldauf, Seljak, \& Senatore}]{Baldauf_2011}
Baldauf, T., Seljak, U., \& Senatore, L. 2011, \bibinfo{title}{Primordial
  non-Gaussianity in the bispectrum of the halo density field,} Journal of
  Cosmology and Astroparticle Physics, 2011, 006,
  \dodoi{10.1088/1475-7516/2011/04/006}

\bibitem[{A. Barreira(2022)Barreira}]{Barreira_2022}
Barreira, A. 2022, \bibinfo{title}{Can we actually constrain fNL using the
  scale-dependent bias effect? An illustration of the impact of galaxy bias
  uncertainties using the BOSS DR12 galaxy power spectrum,} Journal of
  Cosmology and Astroparticle Physics, 2022, 013,
  \dodoi{10.1088/1475-7516/2022/11/013}

\bibitem[{N. Bartolo {et~al.}(2004)Bartolo, Komatsu, Matarrese, \&
  Riotto}]{Bartolo_2004}
Bartolo, N., Komatsu, E., Matarrese, S., \& Riotto, A. 2004,
  \bibinfo{title}{Non-Gaussianity from inflation: theory and observations,}
  Physics Reports, 402, 103–266, \dodoi{10.1016/j.physrep.2004.08.022}

\bibitem[{F. Bernardeau {et~al.}(2002)Bernardeau, Colombi, Gaztanaga, \&
  Scoccimarro}]{Bernardeau:2001qr}
Bernardeau, F., Colombi, S., Gaztanaga, E., \& Scoccimarro, R. 2002,
  \bibinfo{title}{{Large scale structure of the universe and cosmological
  perturbation theory},} Phys. Rept., 367, 1.
\newblock \doarXiv{astro-ph/0112551}

\bibitem[{A. Blanchard {et~al.}(2020)Blanchard {et~al.}}]{Euclid:2019clj}
Blanchard, A., {et~al.} 2020, \bibinfo{title}{{Euclid preparation: VII.
  Forecast validation for galaxy clustering and weak lensing},} Astron.
  Astrophys., 642, A191, \dodoi{10.1051/0004-6361/202038071}

\bibitem[{D. Blas {et~al.}(2011)Blas, Lesgourgues, \& Tram}]{Diego_Blas_2011}
Blas, D., Lesgourgues, J., \& Tram, T. 2011, \bibinfo{title}{The Cosmic Linear
  Anisotropy Solving System (CLASS). Part II: Approximation schemes,} Journal
  of Cosmology and Astroparticle Physics, 2011, 034,
  \dodoi{10.1088/1475-7516/2011/07/034}

\bibitem[{G. Cabass {et~al.}(2022)Cabass, Ivanov, Philcox,
  {et~al.}}]{Cabass:2022}
Cabass, G., Ivanov, M.~M., Philcox, O. H.~E., {et~al.} 2022,
  \bibinfo{title}{{Constraints on multifield inflation from the BOSS galaxy
  survey},} Phys. Rev. D, 106, 043506.
\newblock \doarXiv{2204.01781}

\bibitem[{G. Cabass {et~al.}(2017)Cabass, Pajer, \& Schmidt}]{Cabass_2017}
Cabass, G., Pajer, E., \& Schmidt, F. 2017, \bibinfo{title}{How Gaussian can
  our Universe be?} Journal of Cosmology and Astroparticle Physics, 2017, 003,
  \dodoi{10.1088/1475-7516/2017/01/003}

\bibitem[{M.~S. Cagliari {et~al.}(2025)Cagliari, Barberi-Squarotti, Pardede,
  Castorina, \& D’Amico}]{Cagliari_2025}
Cagliari, M.~S., Barberi-Squarotti, M., Pardede, K., Castorina, E., \&
  D’Amico, G. 2025, \bibinfo{title}{Bispectrum constraints on Primordial
  Non-Gaussianities with the eBOSS DR16 quasars,} Journal of Cosmology and
  Astroparticle Physics, 2025, 043, \dodoi{10.1088/1475-7516/2025/07/043}

\bibitem[{Y. Cao {et~al.}(2018)Cao, Gong, Meng, Xu, Chen, Guo, Li, Liu, Xue,
  Cao, Fu, Zhang, Wang, \& Zhan}]{10.1093/mnras/sty1980}
Cao, Y., Gong, Y., Meng, X.-M., {et~al.} 2018, \bibinfo{title}{Testing
  photometric redshift measurements with filter definition of the Chinese Space
  Station Optical Survey (CSS-OS),} Monthly Notices of the Royal Astronomical
  Society, 480, 2178, \dodoi{10.1093/mnras/sty1980}

\bibitem[{K.~C. Chan {et~al.}(2012)Chan, Scoccimarro, \& Sheth}]{Chan_2012}
Chan, K.~C., Scoccimarro, R., \& Sheth, R.~K. 2012, \bibinfo{title}{Gravity and
  large-scale nonlocal bias,} Physical Review D, 85, 083509,
  \dodoi{10.1103/physrevd.85.083509}

\bibitem[{E. Chaussidon {et~al.}(2025)Chaussidon, Yèche, de~Mattia, Payerne,
  McDonald, Ross, Ahlen, Bianchi, Brooks, Burtin, Claybaugh, de~la Macorra,
  Doel, Ferraro, Font-Ribera, Forero-Romero, Gaztañaga, Gil-Marín, Gontcho,
  Gutierrez, Guy, Honscheid, Howlett, Huterer, Kehoe, Kirkby, Kisner, Kremin,
  Guillou, Levi, Manera, Meisner, Miquel, Moustakas, Newman, Niz,
  Palanque-Delabrouille, Percival, Prada, Pérez-Ràfols, Ravoux, Rossi,
  Sanchez, Schlegel, Schubnell, Seo, Sprayberry, Tarlé, Vargas-Magaña,
  Weaver, Zhao, \& Zou}]{chaussidon2025}
Chaussidon, E., Yèche, C., de~Mattia, A., {et~al.} 2025,
  \bibinfo{title}{Constraining primordial non-Gaussianity with DESI 2024 LRG
  and QSO samples,} Journal of Cosmology and Astroparticle Physics, 2025, 029,
  \dodoi{10.1088/1475-7516/2025/06/029}

\bibitem[{P. Creminelli \& M. Zaldarriaga(2004)Creminelli \&
  Zaldarriaga}]{Paolo_Creminelli_2004}
Creminelli, P., \& Zaldarriaga, M. 2004, \bibinfo{title}{A single-field
  consistency relation for the three-point function,} Journal of Cosmology and
  Astroparticle Physics, 2004, 006, \dodoi{10.1088/1475-7516/2004/10/006}

\bibitem[{D.~J. Croton {et~al.}(2006)Croton, Springel, White, De~Lucia, Frenk,
  Gao, Jenkins, Kauffmann, Navarro, \&
  Yoshida}]{10.1111/j.1365-2966.2005.09675.x}
Croton, D.~J., Springel, V., White, S. D.~M., {et~al.} 2006,
  \bibinfo{title}{The many lives of active galactic nuclei: cooling flows,
  black holes and the luminosities and colours of galaxies,} Monthly Notices of
  the Royal Astronomical Society, 365, 11,
  \dodoi{10.1111/j.1365-2966.2005.09675.x}

\bibitem[{ {CSST Collaboration} {et~al.}(2026){CSST Collaboration}, Gong, Miao,
  Zhan, Li, Shangguan, Li, Liu, Chen, Yuan, Zhou, Liu, Yu, Ji, Qi, Liu, Dai,
  Wang, Zheng, Hao, Dou, Ao, Lin, Zhang, Wang, Sun, Li, Li, Xu, Li, Li, Wu,
  Zhang, Wang, Bai, Cai, Cai, Cao, Chan, Chang, Chen, Chen, Chen, Chen, Cui,
  Dong, Du, Duan, Fan, Fan, Fan, Fan, Fang, Fu, Fu, Fu, Gao, Gu, Gu, Guo, Han,
  Hu, Huang, Ho, Jiang, Jiang, Jing, Kang, Kong, Li, Li, Li, Li, Li, Li, Liao,
  Lin, Liu, Liu, Liu, Liu, Mao, Mao, Meng, Pang, Peng, Peng, Shan, Shen, Shen,
  Shen, Shi, Shi, Tan, Tian, Wang, Wang, Wang, Wang, Wu, Wu, Wu, Xu, Xue, Xue,
  Yang, Yang, Yao, Yuan, Yuan, Zhang, Zhang, Zhang, Zhang, Zhang, Zhao, Zhao,
  Zhong, Zhong, Zhou, Zhu, \& Zu}]{2026}
{CSST Collaboration}, Gong, Y., Miao, H., {et~al.} 2026,
  \bibinfo{title}{Introduction to the Chinese Space Station Survey Telescope
  (CSST),} Science China Physics, Mechanics \& Astronomy, 69, 239501,
  \dodoi{10.1007/s11433-025-2809-0}

\bibitem[{G. De~Lucia \& J. Blaizot(2007)De~Lucia \&
  Blaizot}]{10.1111/j.1365-2966.2006.11287.x}
De~Lucia, G., \& Blaizot, J. 2007, \bibinfo{title}{The hierarchical formation
  of the brightest cluster galaxies,} Monthly Notices of the Royal Astronomical
  Society, 375, 2, \dodoi{10.1111/j.1365-2966.2006.11287.x}

\bibitem[{F. Deng {et~al.}(2022)Deng, Gong, Wang, Dong, Cao, \&
  Chen}]{10.1093/mnras/stac2185}
Deng, F., Gong, Y., Wang, Y., {et~al.} 2022, \bibinfo{title}{Forecasting the
  cross-correlation of the CSST galaxy survey with the FAST H i Intensity
  Map,} Monthly Notices of the Royal Astronomical Society, 515, 5894,
  \dodoi{10.1093/mnras/stac2185}

\bibitem[{S. Dodelson \& F. Schmidt(2020)Dodelson \& Schmidt}]{Dodelson2020}
Dodelson, S., \& Schmidt, F. 2020, Modern Cosmology, 2nd edn. (Academic Press)

\bibitem[{T. Eifler {et~al.}(2021)Eifler, Miyatake, Krause, Heinrich, Miranda,
  Hirata, Xu, Hemmati, Simet, Capak, Choi, Doré, Doux, Fang, Hounsell, Huff,
  Huang, Jarvis, Kruk, Masters, Rozo, Scolnic, Spergel, Troxel,
  von der Linden, Wang, Weinberg, Wenzl, \& Wu}]{Eifler_2021}
Eifler, T., Miyatake, H., Krause, E., {et~al.} 2021, \bibinfo{title}{Cosmology
  with the {Roman Space Telescope} – multiprobe strategies,} Monthly Notices
  of the Royal Astronomical Society, 507, 1746–1761,
  \dodoi{10.1093/mnras/stab1762}

\bibitem[{ {Euclid Collaboration} {et~al.}(2026){Euclid Collaboration}, Linde,
  Moradinezhad~Dizgah, Parimbelli, {et~al.}}]{Euclid:2026_Linde}
{Euclid Collaboration}, Linde, D., Moradinezhad~Dizgah, A., Parimbelli, G.,
  {et~al.} 2026, \bibinfo{title}{{Euclid preparation. Testing multi-field
  inflation with galaxy power spectrum and bispectrum},} Astron. Astrophys.
  (submitted).
\newblock \doarXiv{2605.21436}

\bibitem[{H.~A. Feldman {et~al.}(1994)Feldman, Kaiser, \&
  Peacock}]{Feldman_1994}
Feldman, H.~A., Kaiser, N., \& Peacock, J.~A. 1994,
  \bibinfo{title}{Power-spectrum analysis of three-dimensional redshift
  surveys,} The Astrophysical Journal, 426, 23, \dodoi{10.1086/174036}

\bibitem[{A. Font-Ribera {et~al.}(2014)Font-Ribera
  {et~al.}}]{Font-Ribera:2013rwa}
Font-Ribera, A., {et~al.} 2014, \bibinfo{title}{{DESI and other Dark Energy
  experiments in the era of neutrino mass measurements},} JCAP, 05, 023,
  \dodoi{10.1088/1475-7516/2014/05/023}

\bibitem[{D. Foreman-Mackey {et~al.}(2013)Foreman-Mackey, Hogg, Lang, \&
  Goodman}]{Foreman_Mackey_2013}
Foreman-Mackey, D., Hogg, D.~W., Lang, D., \& Goodman, J. 2013,
  \bibinfo{title}{{emcee}: The MCMC Hammer,} Publications of the Astronomical
  Society of the Pacific, 125, 306–312, \dodoi{10.1086/670067}

\bibitem[{H. Gil-Mar{\'i}n {et~al.}(2017)Gil-Mar{\'i}n
  {et~al.}}]{Gil-Marin:2016}
Gil-Mar{\'i}n, H., {et~al.} 2017, \bibinfo{title}{{The clustering of galaxies
  in the SDSS-III Baryon Oscillation Spectroscopic Survey: RSD measurement from
  the power spectrum and bispectrum of the DR12 BOSS galaxies},} Mon. Not. Roy.
  Astron. Soc., 465, 1757.
\newblock \doarXiv{1606.00439}

\bibitem[{Y. Gong {et~al.}(2019)Gong, Liu, Cao, Chen, Fan, Li, Li, Li, Zhang,
  \& Zhan}]{Gong_2019}
Gong, Y., Liu, X., Cao, Y., {et~al.} 2019, \bibinfo{title}{Cosmology from the
  Chinese Space Station Optical Survey (CSS-OS),} The Astrophysical Journal,
  883, 203, \dodoi{10.3847/1538-4357/ab391e}

\bibitem[{Y. Gong {et~al.}(2025)Gong, Miao, Zhou, Xiong, Song, Jiang, Wang,
  Yan, Wu, Deng, Chen, Fan, Jing, Yang, \& Zhan}]{Gong_2025}
Gong, Y., Miao, H., Zhou, X., {et~al.} 2025, \bibinfo{title}{Future cosmology:
  New physics and opportunity from the China Space Station Telescope (CSST),}
  Science China Physics, Mechanics \& Astronomy, 68, 280402,
  \dodoi{10.1007/s11433-025-2646-2}

\bibitem[{Q. Guo {et~al.}(2011)Guo, White, Boylan-Kolchin, De~Lucia, Kauffmann,
  Lemson, Li, Springel, \& Weinmann}]{10.1111/j.1365-2966.2010.18114.x}
Guo, Q., White, S., Boylan-Kolchin, M., {et~al.} 2011, \bibinfo{title}{From
  dwarf spheroidals to cD galaxies: simulating the galaxy population in a ΛCDM
  cosmology,} Monthly Notices of the Royal Astronomical Society, 413, 101,
  \dodoi{10.1111/j.1365-2966.2010.18114.x}

\bibitem[{J. Han {et~al.}(2025)Han, Li, Jiang, Chen, Wang, Wei, He, He, Zhang,
  Liu, Cui, Gu, Guo, Jing, Kang, Li, Luo, Luo, Pei, Qiu, Tan, Xie, Yang, Yu,
  Yu, \& Zhou}]{han2025jiutiansimulationscsstextragalactic}
Han, J., Li, M., Jiang, W., {et~al.} 2025, \bibinfo{title}{The Jiutian
  simulations for the CSST extra-galactic surveys,} Science China Physics,
  Mechanics \& Astronomy, 68, 109511, \dodoi{10.1007/s11433-025-2712-1}

\bibitem[{B.~M.~B. Henriques {et~al.}(2015)Henriques, White, Thomas, Angulo,
  Guo, Lemson, Springel, \& Overzier}]{10.1093/mnras/stv705}
Henriques, B. M.~B., White, S. D.~M., Thomas, P.~A., {et~al.} 2015,
  \bibinfo{title}{Galaxy formation in the Planck cosmology – I. Matching the
  observed evolution of star formation rates, colours and stellar masses,}
  Monthly Notices of the Royal Astronomical Society, 451, 2663,
  \dodoi{10.1093/mnras/stv705}

\bibitem[{J. Maldacena(2003)Maldacena}]{Maldacena_2003}
Maldacena, J. 2003, \bibinfo{title}{Non-gaussian features of primordial
  fluctuations in single field inflationary models,} Journal of High Energy
  Physics, 2003, 013, \dodoi{10.1088/1126-6708/2003/05/013}

\bibitem[{H. Miao {et~al.}(2024)Miao, Gong, Chen, Huang, Li, \&
  Zhan}]{miao2024forecastingbaomeasurementscsst}
Miao, H., Gong, Y., Chen, X., {et~al.} 2024, \bibinfo{title}{Forecasting the
  BAO Measurements of the CSST galaxy and AGN Spectroscopic Surveys,} Monthly
  Notices of the Royal Astronomical Society, 531, 3991,
  \dodoi{10.1093/mnras/stae1370}

\bibitem[{K.~G. Noeske {et~al.}(2007)Noeske, Weiner, Faber, Papovich, Koo,
  Somerville, {et~al.}}]{Noeske2007}
Noeske, K.~G., Weiner, B.~J., Faber, S.~M., {et~al.} 2007, \bibinfo{title}{Star
  Formation in {AEGIS} Field Galaxies since {$z=1.1$}: The Dominance of
  Gradually Declining Star Formation, and the Main Sequence of Star-forming
  Galaxies,} The Astrophysical Journal, 660, L43, \dodoi{10.1086/517926}

\bibitem[{W. Pei {et~al.}(2024)Pei, Guo, Li, Wang, Han, Hu, Su, Gao, Wang, Luo,
  \& Wei}]{Pei2024}
Pei, W., Guo, Q., Li, M., {et~al.} 2024, \bibinfo{title}{Simulating emission
  line galaxies for the next generation of large-scale structure surveys,}
  Monthly Notices of the Royal Astronomical Society, 529, 4958,
  \dodoi{10.1093/mnras/stae866}

\bibitem[{ {Planck Collaboration} {et~al.}(2020{\natexlab{a}}){Planck
  Collaboration}, Aghanim, Akrami, Ashdown, {et~al.}}]{Planck2020}
{Planck Collaboration}, Aghanim, N., Akrami, Y., Ashdown, M., {et~al.}
  2020{\natexlab{a}}, \bibinfo{title}{Planck 2018 results. VI. Cosmological
  parameters,} Astronomy \& Astrophysics, 641, A6

\bibitem[{ {Planck Collaboration} {et~al.}(2020{\natexlab{b}}){Planck
  Collaboration}, Akrami, Arroja, Ashdown, Aumont, Baccigalupi, Ballardini,
  Banday, Barreiro, Bartolo, Basak, Benabed, Bernard, Bersanelli, Bielewicz,
  Bond, Borrill, Bouchet, Bucher, Burigana, Butler, Calabrese, Cardoso,
  Casaponsa, Challinor, Chiang, Colombo, Combet, Crill, Cuttaia, de~Bernardis,
  de~Rosa, de~Zotti, Delabrouille, Delouis, Valentino, Diego, Doré, Douspis,
  Ducout, Dupac, Dusini, Efstathiou, Elsner, Enßlin, Eriksen, Fantaye,
  Fergusson, Fernandez-Cobos, Finelli, Frailis, Fraisse, Franceschi, Frolov,
  Galeotta, Ganga, Génova-Santos, Gerbino, González-Nuevo, Górski, Gratton,
  Gruppuso, Gudmundsson, Hamann, Handley, Hansen, Herranz, Hivon, Huang, Jaffe,
  Jones, Jung, Keihänen, Keskitalo, Kiiveri, Kim, Krachmalnicoff, Kunz,
  Kurki-Suonio, Lamarre, Lasenby, Lattanzi, Lawrence, Jeune, Levrier, Lewis,
  Liguori, Lilje, Lindholm, López-Caniego, Ma, Macías-Pérez, Maggio, Maino,
  Mandolesi, Marcos-Caballero, Maris, Martin, Martínez-González, Matarrese,
  Mauri, McEwen, Meerburg, Meinhold, Melchiorri, Mennella, Migliaccio,
  Miville-Deschênes, Molinari, Moneti, Montier, Morgante, Moss, Münchmeyer,
  Natoli, Oppizzi, Pagano, Paoletti, Partridge, Patanchon, Perrotta, Pettorino,
  Piacentini, Polenta, Puget, Rachen, Racine, Reinecke, Remazeilles, Renzi,
  Rocha, Rubiño-Martín, Ruiz-Granados, Salvati, Savelainen, Scott, Shellard,
  Shiraishi, Sirignano, Sirri, Smith, Spencer, Stanco, Sunyaev, Suur-Uski,
  Tauber, Tavagnacco, Tenti, Toffolatti, Tomasi, Trombetti, Valiviita, Tent,
  Vielva, Villa, Vittorio, Wandelt, Wehus, Zacchei, \&
  Zonca}]{planckcollaboration2019planck2018resultsix}
{Planck Collaboration}, Akrami, Y., Arroja, F., {et~al.} 2020{\natexlab{b}},
  \bibinfo{title}{Planck 2018 results. IX. Constraints on primordial
  non-Gaussianity,} Astronomy \& Astrophysics, 641, A9,
  \dodoi{10.1051/0004-6361/201935891}

\bibitem[{R. Scoccimarro(2000)Scoccimarro}]{Scoccimarro_2000}
Scoccimarro, R. 2000, \bibinfo{title}{The Bispectrum: From Theory to
  Observations,} The Astrophysical Journal, 544, 597, \dodoi{10.1086/317248}

\bibitem[{R. Scoccimarro {et~al.}(1998)Scoccimarro, Colombi, Fry, Couchman,
  Luo, \& Bouchet}]{Scoccimarro:1997st}
Scoccimarro, R., Colombi, S., Fry, J.~N., {et~al.} 1998,
  \bibinfo{title}{{Nonlinear Evolution of the Bispectrum of Cosmological
  Perturbations},} Astrophys. J., 496, 586.
\newblock \doarXiv{astro-ph/9704075}

\bibitem[{E. Sefusatti {et~al.}(2006)Sefusatti, Crocce, Pueblas, \&
  Scoccimarro}]{Sefusatti:2006eu}
Sefusatti, E., Crocce, M., Pueblas, S., \& Scoccimarro, R. 2006,
  \bibinfo{title}{{Cosmology and the bispectrum},} Phys. Rev. D, 74, 023522.
\newblock \doarXiv{astro-ph/0604505}

\bibitem[{U. Seljak(2009)Seljak}]{PhysRevLett.102.021302}
Seljak, U. 2009, \bibinfo{title}{Extracting Primordial Non-Gaussianity without
  Cosmic Variance,} Phys. Rev. Lett., 102, 021302,
  \dodoi{10.1103/PhysRevLett.102.021302}

\bibitem[{L. Senatore \& M. Zaldarriaga(2012)Senatore \&
  Zaldarriaga}]{Senatore_2012}
Senatore, L., \& Zaldarriaga, M. 2012, \bibinfo{title}{The effective field
  theory of multifield inflation,} Journal of High Energy Physics, 2012, 024,
  \dodoi{10.1007/jhep04(2012)024}

\bibitem[{A. Smith {et~al.}(2022{\natexlab{a}})Smith, Cole, Grove, Norberg, \&
  Zarrouk}]{Smith_2022a}
Smith, A., Cole, S., Grove, C., Norberg, P., \& Zarrouk, P. 2022{\natexlab{a}},
  \bibinfo{title}{Solving small-scale clustering problems in approximate
  light-cone mocks,} Monthly Notices of the Royal Astronomical Society, 516,
  1062–1071, \dodoi{10.1093/mnras/stac2219}

\bibitem[{A. Smith {et~al.}(2022{\natexlab{b}})Smith, Cole, Grove, Norberg, \&
  Zarrouk}]{Smith_2022b}
Smith, A., Cole, S., Grove, C., Norberg, P., \& Zarrouk, P. 2022{\natexlab{b}},
  \bibinfo{title}{A light-cone catalogue from the Millennium-XXL simulation:
  improved spatial interpolation and colour distributions for the DESI BGS,}
  Monthly Notices of the Royal Astronomical Society, 516, 4529–4542,
  \dodoi{10.1093/mnras/stac2519}

\bibitem[{Y. Song {et~al.}(2024)Song, Xiong, Gong, Deng, Chan, Chen, Guo, Li,
  Li, Liu, Luo, Pei, \& Wei}]{song2024}
Song, Y., Xiong, Q., Gong, Y., {et~al.} 2024, \bibinfo{title}{Cosmological
  Prediction of the Void and Galaxy Clustering Measurements in the {CSST}
  Spectroscopic Survey,} The Astrophysical Journal, 976, 244,
  \dodoi{10.3847/1538-4357/ad8de9}

\bibitem[{D. Spergel {et~al.}(2015)Spergel, Gehrels, Baltay, Bennett,
  Breckinridge, Donahue, Dressler, Gaudi, Greene, Guyon, Hirata, Kalirai,
  Kasdin, Macintosh, Moos, Perlmutter, Postman, Rauscher, Rhodes, Wang,
  Weinberg, Benford, Hudson, Jeong, Mellier, Traub, Yamada, Capak, Colbert,
  Masters, Penny, Savransky, Stern, Zimmerman, Barry, Bartusek, Carpenter,
  Cheng, Content, Dekens, Demers, Grady, Jackson, Kuan, Kruk, Melton, Nemati,
  Parvin, Poberezhskiy, Peddie, Ruffa, Wallace, Whipple, Wollack, \&
  Zhao}]{spergel2015widefieldinfrarredsurveytelescopeastrophysics}
Spergel, D., Gehrels, N., Baltay, C., {et~al.} 2015, Wide-Field InfrarRed
  Survey Telescope-Astrophysics Focused Telescope Assets WFIRST-AFTA 2015
  Report, \doarXiv{1503.03757}

\bibitem[{V. Springel(2005)Springel}]{Springel2005}
Springel, V. 2005, \bibinfo{title}{The cosmological simulation code gadget-2,}
  Monthly Notices of the Royal Astronomical Society, 364, 1105,
  \dodoi{10.1111/j.1365-2966.2005.09655.x}

\bibitem[{V. Springel {et~al.}(2001)Springel, Yoshida, \&
  White}]{SPRINGEL200179}
Springel, V., Yoshida, N., \& White, S.~D. 2001, \bibinfo{title}{GADGET: a code
  for collisionless and gasdynamical cosmological simulations,} New Astronomy,
  6, 79, \dodoi{10.1016/S1384-1076(01)00042-2}

\bibitem[{N.~S. Sugiyama {et~al.}(2019)Sugiyama, Saito, Beutler, \&
  Seo}]{Sugiyama_2018}
Sugiyama, N.~S., Saito, S., Beutler, F., \& Seo, H.-J. 2019, \bibinfo{title}{A
  complete FFT-based decomposition formalism for the redshift-space
  bispectrum,} Monthly Notices of the Royal Astronomical Society, 484,
  364–384, \dodoi{10.1093/mnras/sty3249}

\bibitem[{N.~S. Sugiyama {et~al.}(2020)Sugiyama, Saito, Beutler, \&
  Seo}]{Sugiyama2020MNRAS}
Sugiyama, N.~S., Saito, S., Beutler, F., \& Seo, H.-J. 2020,
  \bibinfo{title}{Perturbation theory approach to predict the covariance
  matrices of the galaxy power spectrum and bispectrum in redshift space,}
  Monthly Notices of the Royal Astronomical Society, 497, 1684,
  \dodoi{10.1093/mnras/staa1940}

\bibitem[{M. Tellarini {et~al.}(2016)Tellarini, Ross, Tasinato, \&
  Wands}]{Tellarini_2016}
Tellarini, M., Ross, A.~J., Tasinato, G., \& Wands, D. 2016,
  \bibinfo{title}{Galaxy bispectrum, primordial non-Gaussianity and redshift
  space distortions,} Journal of Cosmology and Astroparticle Physics, 2016,
  014, \dodoi{10.1088/1475-7516/2016/06/014}

\bibitem[{D.~A. Varshalovich {et~al.}(1988)Varshalovich, Moskalev, \&
  Khersonskii}]{doi:10.1142/0270}
Varshalovich, D.~A., Moskalev, A.~N., \& Khersonskii, V.~K. 1988, Quantum
  Theory of Angular Momentum (WORLD SCIENTIFIC), \dodoi{10.1142/0270}

\bibitem[{M.~S. Wang {et~al.}(2023{\natexlab{a}})Wang, Beutler, \&
  Sugiyama}]{Wang:2023a}
Wang, M.~S., Beutler, F., \& Sugiyama, N.~S. 2023{\natexlab{a}},
  \bibinfo{title}{{Triumvirate: A Python/C++ package for three-point clustering
  measurements},} J.~Open~Source~Softw., 8, 5571, \dodoi{10.21105/joss.05571}

\bibitem[{M.~S. Wang {et~al.}(2023{\natexlab{b}})Wang, Beutler, \&
  Sugiyama}]{Wang:2023b}
Wang, M.~S., Beutler, F., \& Sugiyama, N.~S. 2023{\natexlab{b}}, {Triumvirate:
  A Python/C++ package for three-point clustering measurements}, 0.3.0 Zenodo,
  \dodoi{10.5281/zenodo.10072129}

\bibitem[{S. Weinberg(2008)Weinberg}]{Weinberg2008}
Weinberg, S. 2008, Cosmology (Oxford University Press)

\bibitem[{H.-R. {Yu} {et~al.}(2026){Yu}, {Chen}, {Xu}, {Sheng}, {Han}, {Jing},
  \& {Cui}}]{2026SCPMA..6969511Y}
{Yu}, H.-R., {Chen}, B.-H., {Xu}, K., {et~al.} 2026, \bibinfo{title}{{CUBE2: A
  parallel N-body simulation code for scalability, accuracy, and memory
  efficiency},} Science China Physics, Mechanics, and Astronomy, 69, 269511,
  \dodoi{10.1007/s11433-025-2926-0}

\bibitem[{H. Zhan(2011)Zhan}]{132011-961}
Zhan, H. 2011, \bibinfo{title}{Consideration for a large-scale multi-color
  imaging and slitless spectroscopy survey on the Chinese space station and its
  application in dark energy research,} SCIENTIA SINICA Physica, Mechanica \&
  Astronomica, 41, 1441, \dodoi{10.1360/132011-961}

\bibitem[{H. Zhan(2021)Zhan}]{TB-2021-0016}
Zhan, H. 2021, \bibinfo{title}{The wide-field multiband imaging and slitless
  spectroscopy survey to be carried out by the Survey Space Telescope of China
  Manned Space Program,} Chinese Science Bulletin, 66, 1290,
  \dodoi{10.1360/TB-2021-0016}

\end{thebibliography}
\bibliographystyle{aasjournalv7}

\end{document}